\documentclass[fleqn,10pt]{wlscirep}
\usepackage[utf8]{inputenc}
\usepackage[T1]{fontenc}
\usepackage{lmodern}
\usepackage{amsmath}
\usepackage{amsthm}
\usepackage{amsfonts}
\usepackage{amssymb}
\usepackage{amsmath,amscd}
\usepackage{physics}
\usepackage{dsfont}
\usepackage{empheq}
\usepackage{graphicx}
\usepackage{subfig}
\usepackage{tabularx}
\usepackage{nicefrac}
\usepackage{xcolor}
\usepackage{float}
\usepackage{caption}
\usepackage{braket}
\usepackage{algorithm}
\usepackage{algpseudocode}
\usepackage{booktabs}
\usepackage[toc]{appendix}
\usepackage{multirow}
\usepackage{subcaption}
\usepackage[export]{adjustbox}

\DeclareMathOperator*{\argmin}{arg\,min}

\title{Leveraging Analog Quantum Computing with Neutral Atoms for Solvent Configuration Prediction in Drug Discovery}

\author[1, +]{Mauro D'Arcangelo}
\author[2, +]{Daniele Loco}
\author[1]{Fresnel team}
\author[2,3,4]{Nicolaï Gouraud}
\author[2]{Stanislas Angebault}
\author[2]{Jules Sueiro}

\author[3]{Pierre Monmarché}
\author[2]{Jérôme Forêt}
\author[1]{Louis-Paul Henry}
\author[1,*]{Loïc Henriet}
\author[2,4,*]{Jean-Philip Piquemal}

\affil[1]{Pasqal, 7 Rue Léonard de Vinci, 91300 Massy, France}
\affil[2]{Qubit Pharmaceuticals, Advanced Research Department, 24 rue du Faubourg Saint-Jacques, 75014 Paris, France}
\affil[3]{Sorbonne Université, Laboratoire Jacques-Louis Lions, UMR 7589 CNRS, 75005, Paris, France}
\affil[4]{Sorbonne Université, Laboratoire de Chimie Théorique, UMR 7616 CNRS, 75005, Paris, France}
\affil[*]{loic.henriet@pasqal.com, jean-philip.piquemal@sorbonne-universite.fr}

\affil[+]{these authors contributed equally to this work}


\begin{abstract}
   We introduce quantum algorithms able to sample equilibrium water solvent molecules configurations within proteins thanks to analog quantum computing. To do so, we combine a quantum placement strategy to the 3D Reference Interaction Site Model (3D-RISM), an approach capable of predicting continuous solvent distributions. The intrinsic quantum nature of such coupling guarantees molecules not to be placed too close to each other, a constraint usually imposed by hand in classical approaches. We present first a full quantum adiabatic evolution model that uses a local Rydberg Hamiltonian to cast the general problem into an anti-ferromagnetic Ising model.  Its solution, an NP-hard problem in classical computing, is embodied into a Rydberg atom array Quantum Processing Unit (QPU). Following a classical emulator implementation, a QPU portage allows to experimentally validate the algorithm performances on an actual quantum computer. As a perspective of use on next generation devices, we emulate a second hybrid quantum-classical version of the algorithm. Such a variational quantum approach (VQA) uses a classical Bayesian minimization routine to find the optimal laser parameters. Overall, these Quantum-3D-RISM (Q-3D-RISM) algorithms open a new route towards the application of analog quantum computing in molecular modelling and drug design.
    \end{abstract} 

\usepackage{xr}
\makeatletter

\newcommand*{\addFileDependency}[1]{
\typeout{(#1)}
\@addtofilelist{#1}
\IfFileExists{#1}{}{\typeout{No file #1.}}
}\makeatother

\newcommand*{\myexternaldocument}[1]{%
\externaldocument{#1}%
\addFileDependency{#1.tex}%
\addFileDependency{#1.aux}%
}

\myexternaldocument{suppinfo}

\begin{document}

\flushbottom
\maketitle

The direct manipulation of quantum systems to perform Quantum Computations has become an intense field of interdisciplinary research aiming to apply quantum computing to various fields ranging from theoretical chemistry and many-body physics to material sciences and drug design. For example, such technologies are believed to be ultimately able to improve the accuracy of Quantum Chemistry methods ~\cite{ChemRev_AG2019,RevModPhys.92.015003,feniou2023overlapadaptvqe,feniou2023greedy} through algorithms exponentially faster than classical ones.~\cite{ChemRev_AG2019,RevModPhys.92.015003,RevModPhys.94.015004}. If no widely accepted proof of exponential advantage for quantum computing calculations has been yet provided~\cite{Chan_QC}, in practice, a polynomial advantage could still provide more accurate simulations at hand in near-term quantum devices~\cite{ChemRev_AG2019,Chan_QC,RevModPhys.94.015004,Cerezo2021}. More pragmatically, a lower energy consumption of quantum devices is expected to be more rapidly envisioned.

Reasons for this present situation are linked to near-term quantum technologies limitations in term of qubit count and to the absence of fully error-corrected gate-based machines leading to high levels of noise ~\cite{RevModPhys.92.015003,Cerezo2021,PhysRevA.107.012416}. In that context, analog quantum computing is an alternative promising strategy \cite{henriet2020quantum,scholl2021quantum}.

Interestingly, on the software side, many groups have contributed to develop algorithms that are expected to scale efficiently with the size of chemical systems, once that more advanced hardware will become available~\cite{ChemRev_AG2019,RevModPhys.92.015003}. Such techniques include the Variational Quantum Algorithms (VQAs)~\cite{RevModPhys.94.015004,Cerezo2021}, a family of algorithms allowing to exploit present hardware through coupling with classical optimization. In a nutshell, if one is able to formulate a problem in terms of a mathematical cost function to be minimized/maximized with respect to the parameters of the quantum machine, the search for the best parameters can be outsourced to a classical hardware, the quantum computer being used to compute the cost function, a task usually too expensive to compute classically. The solution to the problem is then given in terms of the best parameters for the quantum evolution. 
Concerning the application of quantum computing to drug design, a necessary tool for accurate drug-target affinity predictions in structure-based Drug Discovery (DD) is related to methods capable of addressing the problem of predicting the hydration sites in protein structures and in any biomolecular systems to a larger extend. Predicting the solvent structure inside important protein pockets thanks to user-friendly and computationally-efficient methods is a determining task to improve the predicting power of numerical simulations. Indeed, it requires to deal with the computational effort of sampling a highly dimensional phase space, generated by the daunting number of degrees of freedom of the combined protein-environment complex system. There are methods such as 3D Reference Interaction Site Model (3D-RISM) that can avoid this effort through the use of continuous distributions~\cite{chandler_optimized_1972,chandler_derivation_1973,chandler_equilibrium_1974,beglov_integral_1997,kovalenko_three-dimensional_2003,roy_biomolecular_2021} based on a statistical mechanics integral equation formalism.  However they lack information about the exact locations of molecules.
In this work, we present the first algorithms, exploiting the analog quantum computing paradigm, to analyze the hydration structures within proteins’
binding sites thanks to the 3D-RISM continuous distribution. Indeed, a poor treatment of the protein-hydration problem is often a limiting factor preventing the access to accurate protein-
ligand affinity predictions, a
crucial step for practitioners in drug discovery. Our strategy is applied to small molecules of interest in drug design for which the effect of solvent water molecules is substantial. The main advantage of this proposed approach is its potential to efficiently sample the distribution of water molecules inside protein cavities.  Therefore, this work presents the design of novel quantum algorithmic strategies in that direction. To do so, we implemented the approach on both PASQAL’s Pulser emulator and real neutral atom quantum computer, which, thanks to its flexible connectivity and ease of implementation of analog operations, has demonstrated the potential to solve combinatorial problems. The sampling of water molecules in biological systems distribution can also be seen as a combinatorial problem, since it corresponds to finding the most probable disposition of solvent molecules given the constraints imposed by both the protein structure and the rest of the solvent atomistic structure.

\textbf{The importance of water placement}. More precisely, the distribution of water molecules impacts the protein's structure and determines its overall shape. Water molecules also mediate the interaction between proteins and other small molecules, the so-called
ligands. The presence of water generally influences the binding capacity of a ligand to a specific protein site~\cite{Samways2021,Bucher2018}, a problem of particular interest for the discovery of new drugs. In practice, water molecules can occupy ligand-protein interaction sites, and clusters of water molecules can deeply affect the thermodynamics of ligand binding modes. Consequently, accounting for protein solvation effects is crucial in structure-based drug design, to reliably model the ligand activity through a rational design of the ligand properties~\cite{Samways2021,Bucher2018}. To a certain extent, the presence of water molecules can be experimentally established using X-ray crystallography, or other techniques. However, experiments can present accuracy limitations due to poor resolution~\cite{Wlodawer2008} and it is then critical to refine the structure using accurate molecular modeling tools.

\begin{figure}[h]
    \centering
    \includegraphics[width=0.7\textwidth]{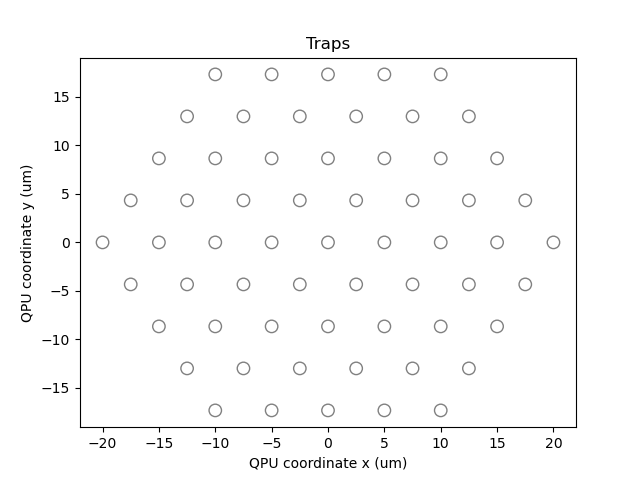}
    \caption{An example of layout of optical traps in a neutral atom QPU. Each circle represents a site were a Rubidium atom can be trapped and used for quantum computations.}
    \label{fig:trap_layout}
\end{figure}

\textbf{Analog quantum computing with neutral atoms.} To address the solvent sampling problem, we propose an approach based on a neutral atom Quantum Processing Unit (QPU) coupled to a 3D-RISM computation. The machine employs arrays of Rubidium atoms arranged in arbitrary 2D configurations defined by a layout of optical traps of the type shown in Fig.~\ref{fig:trap_layout}. Each atom is described by its position $\vec{q}_i$ and its internal state, which can be in general an arbitrary superposition of the ground state $\ket{0}$ and a highly excited Rydberg state $\ket{1}$. The quantum state of the collective $N$-qubit array at time $t$ will be denoted $\ket{\psi_t}$. 

A laser system is tuned close to the resonant frequency of the $\ket{0}$ to $\ket{1}$ transition, so that each atom is effectively a qubit living in a two-dimensional Hilbert space. The programmable time-dependent control fields of the driving laser are the Rabi frequency and detuning, denoted $\Omega(t)$ and $\Delta(t)$ respectively. After evolving for a period of time $T$, the system can be measured in the computational basis by a fluorescence imaging process that will find certain atoms in $\ket{0}$, while the rest are inferred to be in $\ket{1}$\,\cite{henriet2020quantum}.

The term \textit{ground state} will be used in two different contexts. The ground state of a single isolated atom is denoted $\ket{0}$, and the \textit{all-ground state} of a non-interacting $M$-qubit system refers to the state $\ket{\psi_0} := \ket{0} \otimes \ldots \otimes \ket{0} = \ket{0}^{\otimes M}$ which is often the initial state of any quantum algorithm. The ground state of the system coupled with interactions and a driving laser, however, corresponds to the state of minimal energy of the Hamiltonian, which is not necessarily the all-ground state. For instance, the Hamiltonian at $\Omega = 0$ and $\Delta > 0$ can be related to a known graph-theoretical problem. By interpreting each neutral atom as a vertex in a graph, and connecting with an edge vertices situated at a distance closer than a certain threshold dictated by the physics of interactions, the ground state is a solution to the Maximum Independent Set (MIS) problem \cite{max_indep_set_2009}, i.e. the problem of selecting the largest number of vertices in a graph that are not connected to each other directly by an edge.

\textbf{A quantum Ising model formulation of the placement problem.}\label{sec:distance-is-an-ising-model}
To study the water placement problem from a quantum perspective, we place ourselves in the formalim of 3D-RISM. Indeed, this algorithm produces a continuous map of the oxygen atom density inside the protein cavity. We want to translate this continuous information into a discrete set of water molecule positions. In order to do so, we assume that each water molecule randomly oscillates around a definite stable position in the protein cavity. If one tracks over time the oscillating motion of a water molecule and assigns to each unit of space the probability of finding the molecule there, the resulting probability map should resemble a Gaussian of a certain width centered around the stable position. In this ansatz, the 3D-RISM density is interpreted as the sum of all the Gaussians associated to each water molecule in the cavity. Therefore, the joint density is decomposed into several single-particle densities expressing the uncertainty in the position of each water molecule, and the problem of locating the position of water molecules is transposed to the problem of finding the best Gaussian mixture that approximates the 3D-RISM density.

Formally, a protein cavity can be modeled as a connected subset $\mathcal{C} \subset \mathbb{R}^3$ and a 3D-RISM density is a scalar function $g: \mathcal{C} \rightarrow \mathbb{R}$ normalized such that $\int_\mathcal{C} g(\vb{r}) d\vb{r} = 1$. Denote by $\mathcal{G}(\vb*{\mu}, \sigma^2)$ a normalized isotropic Gaussian with mean $\vb*{\mu} \in \mathbb{R}^3$ and variance $\sigma^2$. For practical purposes, we restrict the possible values of the mean of the Gaussians to a finite discrete lattice $Q := \{\vb{q}_1, \ldots, \vb{q}_M\} \subset \mathcal{C}$ rather than the whole continuous space. Assigning a binary variable $n_i \in \{0,1\}$ to each point of $Q$, an arbitrary sum of Gaussians can be written:
\begin{equation}\label{eq:anysog}
    \sum_{i=1}^M \mathcal{G}(\vb{q}_i, \sigma^2) n_i.
\end{equation}
The value of each $n_i$ will then act as a switch that indicates whether or not a Gaussian is placed in position $\vb{q}_i$. We want to find the optimal assignment of $\{n_i\}_{i=1,\ldots,M}$ such that the following $L^2$ norm is minimized:
\begin{equation}\label{eq:l2distance}
    I^2 := \int_\mathcal{C} \left( g(\vb{r}) - \sum_{i=1}^M \mathcal{G}(\vb{q}_i, \sigma^2)(\vb{r}) \ n_i  \right)^2 d\vb{r}.
\end{equation}
Expanding the square, one sees that $I^2$ defines the energy of an Ising model:
\begin{equation}\label{eq:ising_classical}
    I^2 = K - \sum_{i=1}^M \Gamma_i n_i + \sum_{i\neq j=1}^M V_{ij} n_i n_j
\end{equation}
where $K$ is an unimportant constant, and the coefficients of the linear and quadratic terms are given by:
\begin{equation}\label{eq:gamma-local}
    \Gamma_i := 2\int_\mathcal{C} g(\vb{r}) \ \mathcal{G}(\vb{q}_i,\sigma^2)(\vb{r}) \ d\vb{r} - \int_\mathcal{C} \Big( \mathcal{G}(\vb{q}_i,\sigma^2)(\vb{r})\Big)^2 \ d\vb{r}
\end{equation}
and
\begin{equation}
    V_{ij} := \int_\mathcal{C} \mathcal{G}(\vb{q}_i,\sigma^2)(\vb{r}) \  \mathcal{G}(\vb{q}_j,\sigma^2)(\vb{r}) \ d\vb{r}.
\end{equation}
For fixed $\sigma^2$, the coefficients can be computed numerically, but it is helpful to gain further insight into the interaction term $V_{ij}$. When the Gaussians are concentrated far enough from the boundary of $\mathcal{C}$, it is possible to approximate the integral by extending the integration to the whole of $\mathbb{R}^3$. By performing the change of variables $\vb{r} \rightarrow \vb{r}+\vb{q}_i$, the term becomes the convolution of two Gaussians centered in zero:
\begin{equation}
    V_{ij} \simeq \int_{\mathbb{R}^3} \mathcal{G}(\vb{0},\sigma^2)(\vb{r}) \mathcal{G}(\vb{r}_{ij},\sigma^2)(\vb{r}) d\vb{r} = \left[\mathcal{G}(\vb{0},\sigma^2) * \mathcal{G}(\vb{0},\sigma^2)\right] (\vb{r}_{ij})
\end{equation}
where $\vb{r}_{ij} = \vb{q}_j - \vb{q}_i$. It is a known result that the convolution of two Gaussians is itself a Gaussian, therefore one has:
\begin{equation}
    V_{ij} \sim \exp\left(-\alpha |\vb{r}_{ij}|^2\right).
\end{equation}
Equation \eqref{eq:ising_classical} is therefore a classical Ising model with exponentially decaying interactions. In addition, the water placement problem requires two water molecules to be placed at a minimal physical distance from each other. This translates to an extra constraint on the Ising model \eqref{eq:ising_classical} in the form of $n_i = n_j = 1 \implies |\vb{r}_i - \vb{r}_j| > R$ for some $R>0$. If $\{n_1^*, \ldots, n_M^*\}$ is the ground state of the constrained Ising model, the solution to the water placement problem is then defined as: 

\begin{align}
\mathcal{W} &:= \{\vb{q}_i \in Q \mid n_i^* = 1 \} \\
\mathcal{N} &:= \sum_{i=1}^M n_i^*
\end{align}
with $\mathcal{N}$ the number of placed water molecules and $\mathcal{W}$ their positions. This formulation has far-reaching applications to general finite Gaussian mixture models. See Supplementary Information for more details.

The quantum version of an Ising model~\cite{ising_quantum, ising_quantum2} is obtained by replacing the binary variable $n_i$ with the number operator $\hat{n}_i$, whose spectrum is $\{0,1\}$. In terms of Pauli matrices, the number operator is $\hat{n}_i = (\hat{\sigma}_i^z + \mathds{1})/2$. Performing the substitution $n_i \rightarrow \hat{n}_i$, we can establish a direct identification between the classical Ising model \eqref{eq:ising_classical} and the (diagonal) Hamiltonian of a system of interacting spins located in $\vb{q}_1, \ldots, \vb{q}_M$:
\begin{equation}\label{eq:problem_hamiltonian}
    I^2 \rightarrow \hat{I}^2 := - \sum_{i=1}^N \Gamma_i \hat{n}_i +\sum_{i\neq j=1}^{N} V_{ij} \hat{n}_i \hat{n}_j
\end{equation}
which we will be referring to as the \textit{problem Hamiltonian}. If $\mathcal{H}$ denotes the Hilbert space of the quantum system and $\mathcal{B}(\mathcal{H})$ denotes its computational basis (the basis in which the operator in eq.~\eqref{eq:problem_hamiltonian} is diagonal), the ground state of \eqref{eq:problem_hamiltonian} corresponds to the bitstring that minimizes \eqref{eq:ising_classical}, but it is now seen as the computational basis vector $\ket{e^*} \in \mathcal{B}(\mathcal{H})$ that minimizes the expectation value of $\hat{I}^2$:
\begin{equation}\label{eq:bs_best}
    \ket{e^*} := \argmin_{\ket{e} \in \mathcal{B}(\mathcal{H})} \bra{e} \hat{I}^2 \ket{e}.
\end{equation}
The position and number of the water molecules are then given by:
\begin{align}\label{eq:wat_numcoord}
    \mathcal{W} &:= \{\vb{q}_i \in Q \mid \ \bra{e^*}\hat{n}_i \ket{e^*} = 1 \} \\
    \mathcal{N} &:=  \sum_{i=1}^M \bra{e^*} \hat{n}_i \ket{e^*}.\nonumber
\end{align}

\textbf{Solving the quantum Ising problem with a local Rydberg Hamiltonian.} A system of neutral atoms coupled to an optical laser can be crafted in such a way as to evolve according to the following time-dependent Hamiltonian:  
\begin{equation}\label{eq:local_rydberg_hamiltonian}
    \hat{H}(t) = \sum_{i=1}^M \Omega_i(t)\hat{\sigma}^x_i - \sum_{i=1}^M \Delta_i(t) \hat{n}_i +\sum_{i<j=1}^{M} U_{ij} \hat{n}_i \hat{n}_j
\end{equation}
where $\Omega_i(t)$ is the Rabi frequency of the driving laser on qubit $i$, $\Delta_i(t)$ is the detuning of the laser, $U_{ij}$ is the interaction coefficient between Rydberg excitations:
\begin{equation}
    U_{ij} = \frac{C_6}{|\vb{r}_{ij}|^6}
\end{equation}
and $C_6$ is a physical constant. Before describing each term more in detail, one can immediately see the similarity between \eqref{eq:problem_hamiltonian} and \eqref{eq:local_rydberg_hamiltonian}, and consequently why systems of neutral atoms might be particularly suited for solving such a problem. The biggest difference lies in the coefficient of the two-body term: in the Rydberg Hamiltonian $U_{ij}$ decays as a power-law, while in the problem Hamiltonian $V_{ij}$ decays exponentially. However, one should keep in mind that \eqref{eq:problem_hamiltonian} does not directly encode the proximity constraint between neighbouring excitations, which is taken care of precisely by the $r^{-6}$ part of the Rydberg Hamiltonian.

Having control over the Rydberg Hamiltonian \eqref{eq:local_rydberg_hamiltonian}, the solution to the water placement problem can be found by applying the adiabatic theorem~\cite{Farhi2001}. In a nutshell, the strategy is to identify a path $\hat{H}(s)$ in the space of admissible Hamiltonians for some normalized time parameter $s \in [0,1]$ such that the system is initialized in the ground state of $\hat{H}(0)$ (which therefore has to be known and easy to prepare), while the only known property of the ground state of $\hat{H}(1)$ is that it encodes the solution to a hard combinatorial optimization problem. If the evolution is performed slowly enough, then the system remains at all times in the instantaneous ground state of $\hat{H}(s)$ for all $s$, and therefore measurements of the system at $s=1$ give the solution to the optimization problem. 

\begin{figure}[!htbp]
    \centering
    \begin{tabular}{@{}c@{}}
        \includegraphics[width=.7\textwidth]{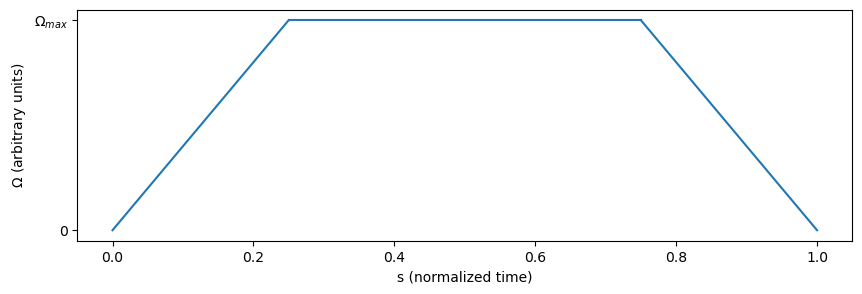} 
    \end{tabular}

\vspace{\floatsep}

    \begin{tabular}{@{}c@{}}
        \includegraphics[width=.7\textwidth]{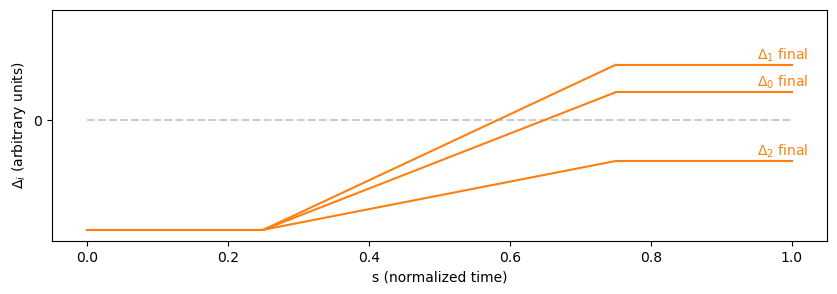} 
    \end{tabular}

    \caption{Schematic example of an adiabatic path parametrized by $\Omega_i(s)$ and $\Delta_i(s)$ in the Rydberg Hamiltonian \eqref{eq:local_rydberg_hamiltonian} for three qubits. The Rabi frequency is kept the same for all qubits, so $\Omega_i(s) = \Omega(s)$ for all $i$, and it vanishes at the extremal points of the path. The Detunings $\Delta_i(s)$ are ramped up from a negative value to some final value related to the one-body terms of the problem Hamiltonian \eqref{eq:ising_classical}, and therefore they will be different for each qubit.}
    \label{fig:adiabatic_pulses}
\end{figure}

We can exploit the quantum adiabatic evolution (QAE) algorithm by choosing $\hat{H}(0)$ to be \eqref{eq:local_rydberg_hamiltonian} with $\Omega_i(0) = 0$, $\Delta_i(0) = -c$ for some large positive constant $c$, so that the ground state of $\hat{H}(0)$ is, to a very good approximation, the all-ground state $\ket{0}^{\otimes M}$. The final Hamiltonian $\hat{H}(1)$ instead is chosen in such a way as to maximize the overlap between its low-energy spectrum and the one of the problem Hamiltonian $\hat{I}^2$. This can be achieved by mapping the one-body terms $\Gamma_i$ in \eqref{eq:ising_classical} to the final detunings $\Delta_i(1)$ in \eqref{eq:local_rydberg_hamiltonian}. A schematic example of $\Omega_i(s)$ and $\Delta_i(s)$ is shown in Fig.~\ref{fig:adiabatic_pulses}.

QAE, however, cannot be implemented on the current generation of neutral atom QPUs because of the lack of local addressing. In other words, instead of the local terms $\Omega_i(t)$ and $\Delta_i(t)$, one only has access to a global control field $\Omega(t)$ and $\Delta(t)$. Nevertheless, the QAE algorithm, described in Algorithm~\ref{alg:HQC},  can be still tested using the Pulser classical emulator\,\cite{pulser}.

\begin{algorithm}[htpb]
\caption{Quantum Adiabatic Evolution (QAE) algorithm that solves the corresponding Ising problem in eq.~\eqref{eq:problem_hamiltonian} using local lasers.\newline
\textbf{inputs}: reference 2D density $g(\vec{r}) : \mathcal{C} \rightarrow \mathbb{R}$ and set of qubits position $Q := \{\vec{q}_i\} \subset \mathcal{C}$, with $\mathcal{C} \subset \mathbb{R}^{2}$. $\Omega$ is set according to the known adiabatic protocol, fitting the three points $[0,\Omega_{Max},0]$ during the whole pulse duration $[0,T]$ and $\Delta$ is computed according to eq.~\eqref{eq:gamma-local}.\newline
\textbf{outputs}: water molecule positions $\mathcal{W}$ and their number $\mathcal{N}$.
}
\label{alg:HQC}
\begin{algorithmic}[1]
\Statex
\Procedure{QAEAlgo}{$g(\vec{r}), \{\vec{q}_i\}$}
    \Statex
    \State $\Delta_i(T) \gets$ \Call{FinalDetuningPerQubit}{$g(\vec{r}), \{\vec{q}_i\}$}
    \State $\Omega(t),\Delta_i(t) \gets$ \Call{BuildAdiabaticPulse}{$\Delta_i(T)$}
    \State $\ket{e^*} \gets $ \Call{QuantumAdiabaticSelection}{$\{\vec{q}_i\}, \Omega(t), \Delta_i(t)$}
    \Comment{Quantum evolution according to the built laser pulse and identification of the best basis state for the given $g(\vec{r})$}
    \Statex
    \State $(\mathcal{W}, \mathcal{N}) \gets$ \Call{PositionsFromState}{$\ket{e^*}$}
    \Comment{$(\mathcal{W}, \mathcal{N})$ are defined in eq.~\eqref{eq:wat_numcoord}}
    \State \textbf{return} $(\mathcal{W}, \mathcal{N})$ \Comment{Best placement of water molecules in, \textit{e.g.}, target protein cavity}
    \Statex
\EndProcedure
\end{algorithmic}
\end{algorithm}

We first tested this approach using synthetic densities. Results can be found in SI, Section~\ref{sec:num-sim-supportQAE}.

\begin{figure}[!htbp]
   \includegraphics[width=0.95\textwidth]{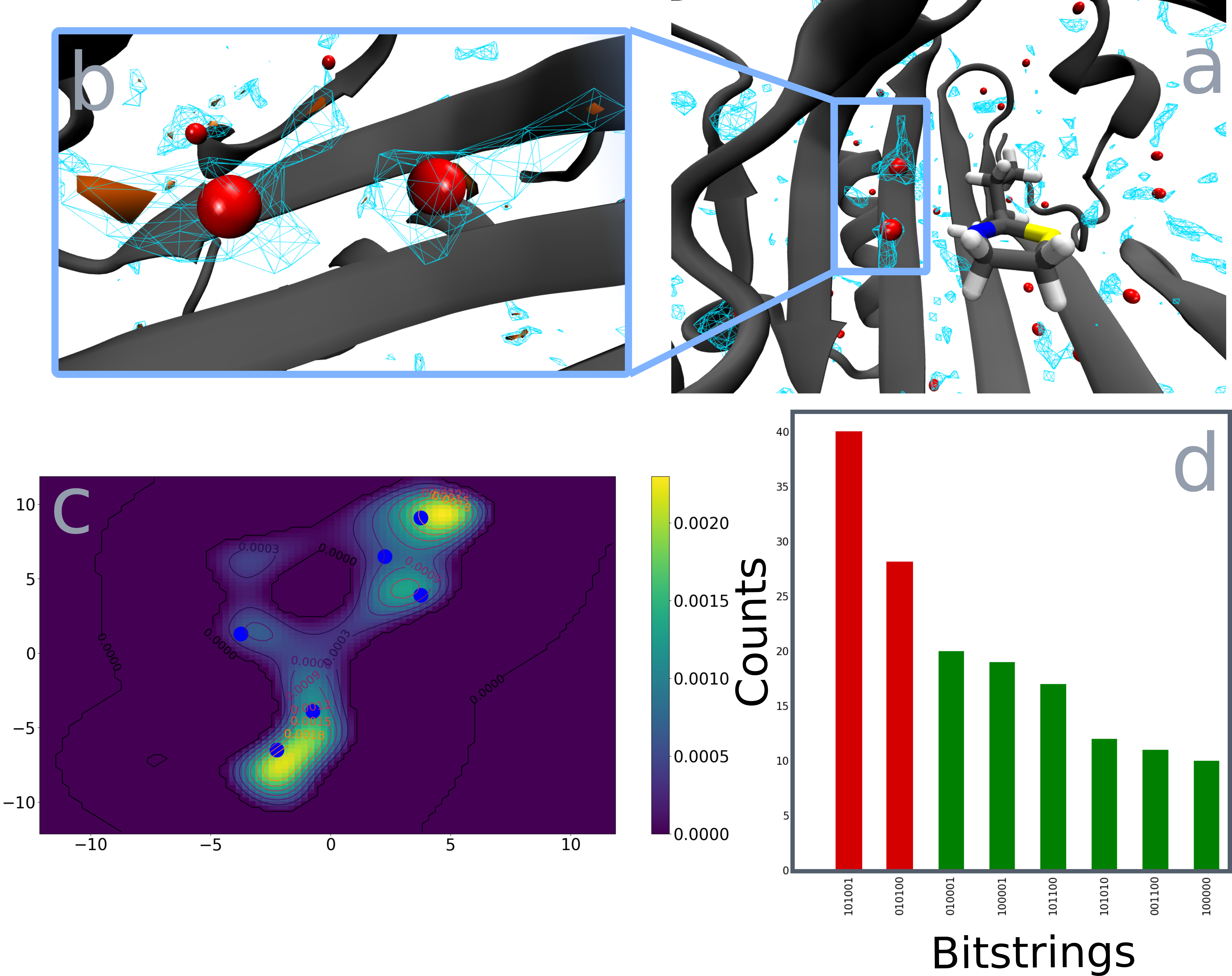}
     \caption{a) View from inside the MUP-I protein pocket, in a complex with the 2-sec-butyl-4,5-dihydrothiazole, in licorice style (PDB entry: 1i06.pdb). The 3D-RISM isodensity with isovalue 5.5 is represented in wireframe style, in cyan; the oxygen of the crystal water molecules placed inside the protein cavity are represented in red. The VMD software is used for the visualization. b) Zoom on the crystal water molecules position: the cyan wireframe 3D-RISM isosurface (isovalue $= 5$) is compared with the orange solid one (isovalue $= 8.5$). c) Representation of the smoothed 3D-RISM density slice, with 6 qubits. d) Histograms reporting the 6-qubit quantum state composition, in terms of basis stats, as obtained from the VQA emulated through Pulser on CPU, without noise, with most sampled bitstring highlighted, corresponding to basis states 101001 (first most sampled) and 010100 (second most sampled).}
    \label{fig:MUP_strct}
\end{figure}

\textbf{Local algorithm emulation using 3D-RISM densities.}
To test the algorithm just presented, we compute the 3D-RISM solvent density within a real protein. We chose the major urinary protein (MUP-I) pocket, where a small ligand, the 2-sec-butyl-4,5-dihydrothiazole, is binded to the protein (see Fig.~\ref{fig:MUP_strct}). This choice is motivated by the fact that the protein-ligand complex structure has been co-crystallized, its complete atomic structure being available from the Protein Data Bank (PDB). Interestingly, the crystal structure presents two structural (i.e. stable) water molecules present in the vicinity of the ligand. Such water molecules are clearly visible in the 3D-RISM density which exhibits high density spots in the same region as the crystal structure (see Fig.~\ref{fig:MUP_strct}). The two water molecules are included in a diffused volume of 3D-RISM density that is identified by a density isovalue surfaces of medium value (see Fig.~\ref{fig:MUP_strct} caption) whereas maximal density values are spotted close to the experimental crystal structures water molecules positions (see orange solid surface in Fig.~\ref{fig:MUP_strct}, b).

\begin{figure}
    \centering
    \includegraphics[width=0.85\textwidth]{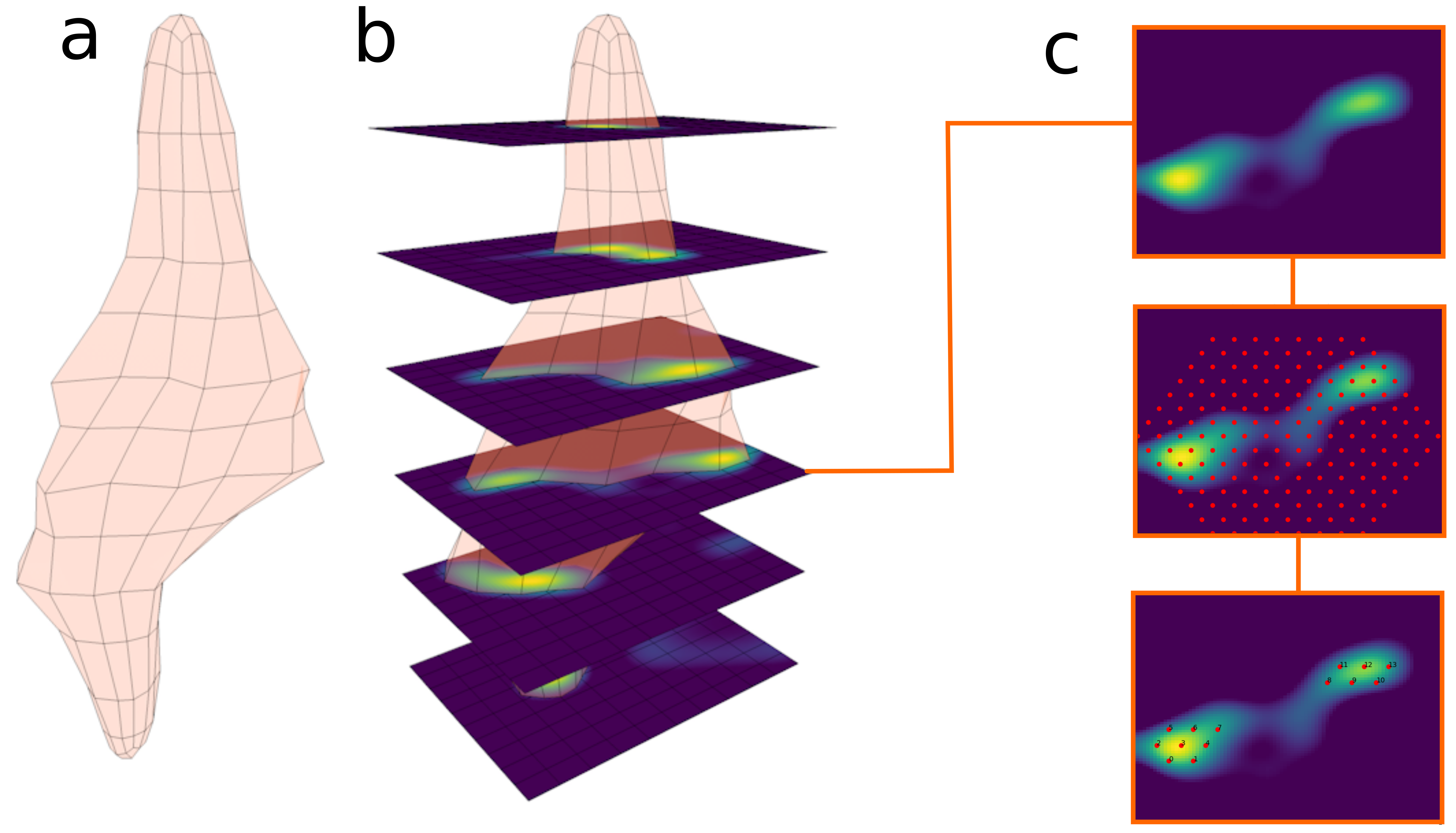}
    \caption{a) Example representation of a 3D-RISM density as a closed isosurface, with a define, constant, density value, as obtained starting from the AmberTools output, discretized on a given chosen grid of point.  b) projection of the 3D-RISM density on 2D planes, namely slices, passing through the cross-section of the density; each slice is a 2D density map, where the 3D-RISM density value discretized on the grid is projected on the plane according to a proximity criterion. c) Process of selection of the qubits array on each 2D density slice: starting from the slice, a regular mesh of traps is disposed uniformly on the density, to then select a limited number of locations where to place qubits according to the local density value; a threshold on the density is set, so that only traps close to a defined density or higher will be occupied.}
    \label{fig:3DRISM_slicing}
\end{figure}

\begin{figure}[!htbp]
    \centering
\begin{tabular}{ll}
    \includegraphics[width=.4\linewidth,valign=t]{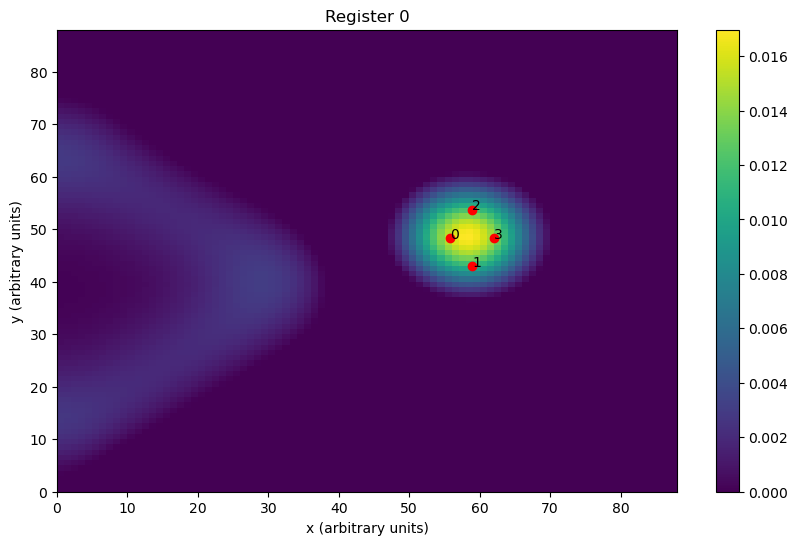} & \includegraphics[width=.4\linewidth,valign=t]{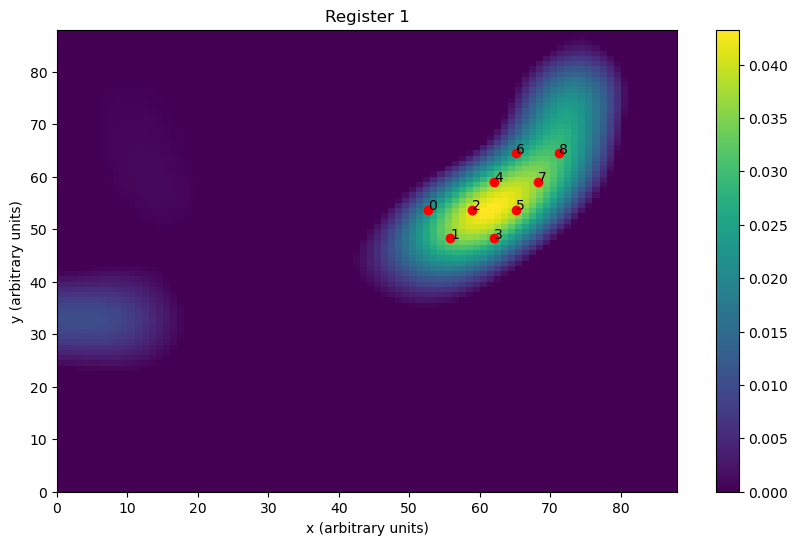} \\ \includegraphics[width=.4\linewidth,valign=t]{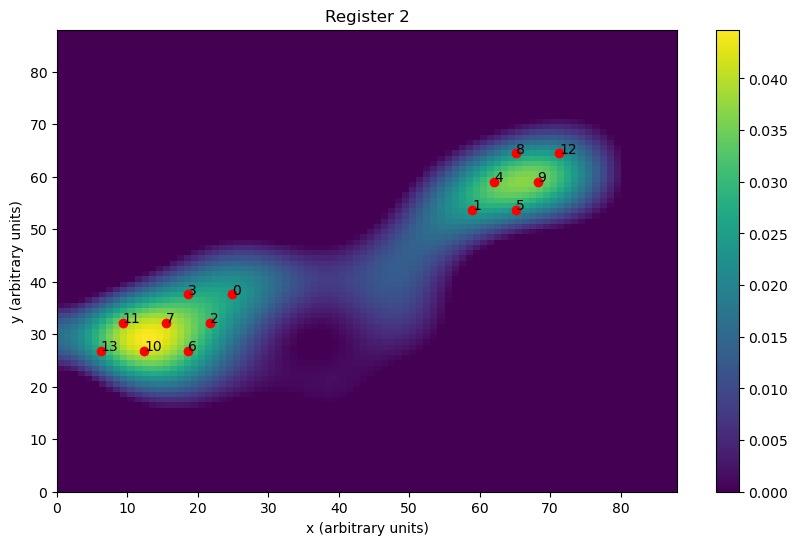} &
    \includegraphics[width=.4\linewidth,valign=t]{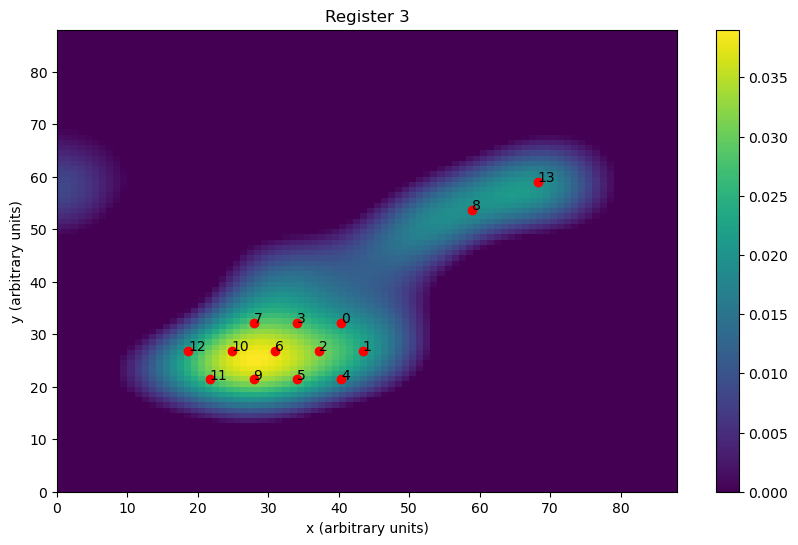} \\ \includegraphics[width=.4\linewidth,valign=t]{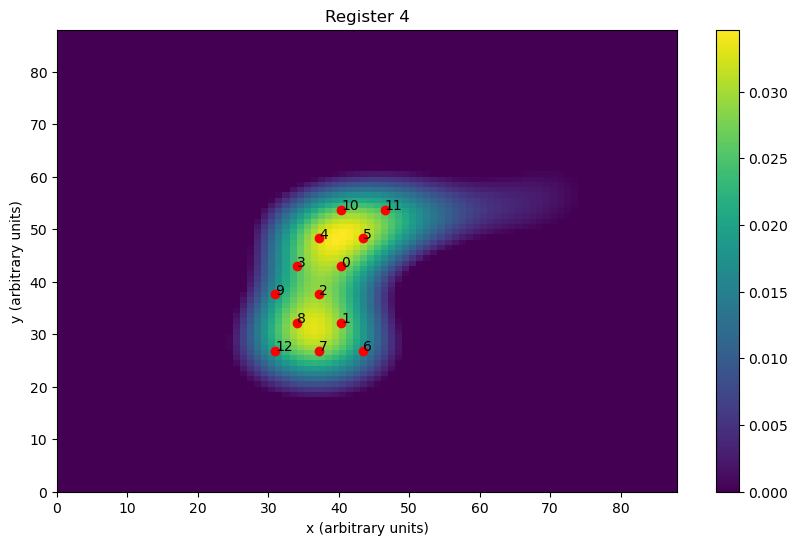} & \includegraphics[width=.4\linewidth,valign=t]{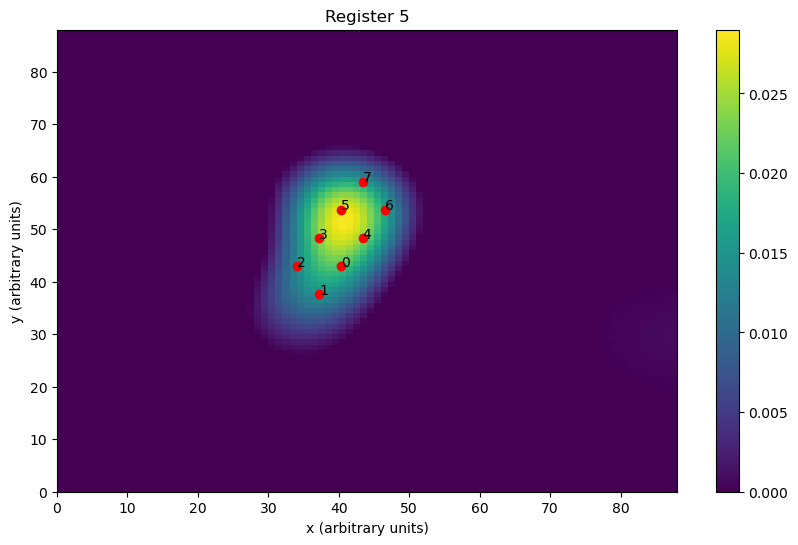}\\
\end{tabular}
\caption{Two-dimensional 3D-RISM density slices and qubit registers used in the local algorithm emulation. The registers vary in size from 4 to 14 qubits. The units displayed in the $x$ and $y$ directions correspond to the 3D-RISM discretization grid, and they are not representative of neither the size of the protein cavity in \AA, nor the size of the qubit registers in $\mu m$. To get an idea of the physical size of the registers, consider the spacing between neighbouring qubits to be fixed at 5 $\mu m$. }
    \label{fig:local_algo_density_qubits}
\end{figure}

\begin{figure}[!htbp]
    \centering
\begin{tabular}{ll}
    \includegraphics[width=.4\linewidth,valign=t]{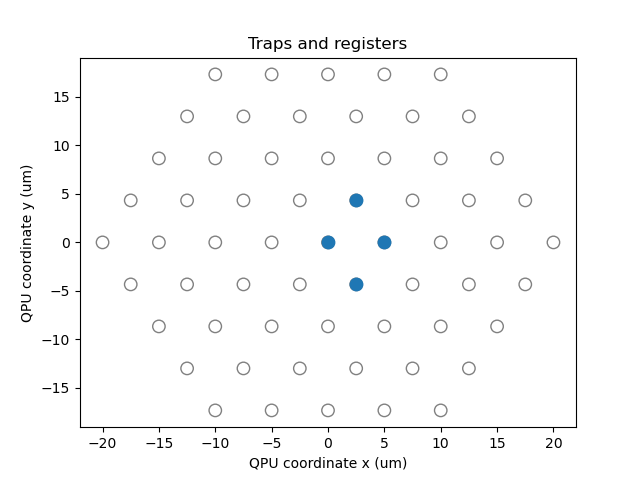} & \includegraphics[width=.4\linewidth,valign=t]{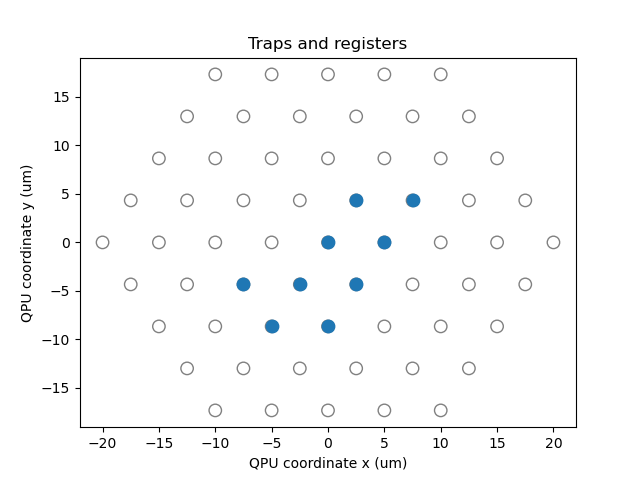} \\ \includegraphics[width=.4\linewidth,valign=t]{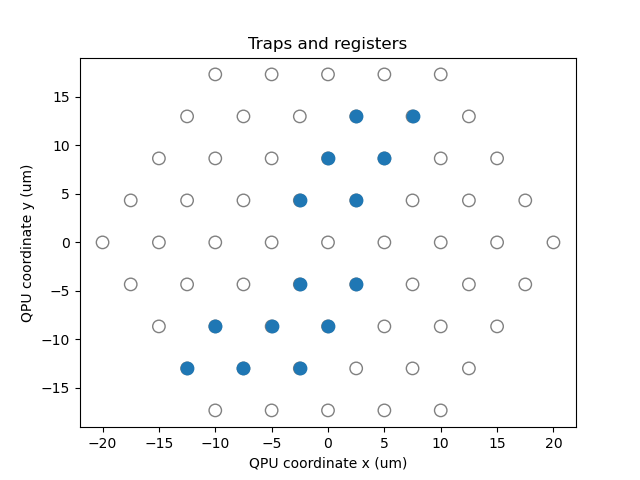} &
    \includegraphics[width=.4\linewidth,valign=t]{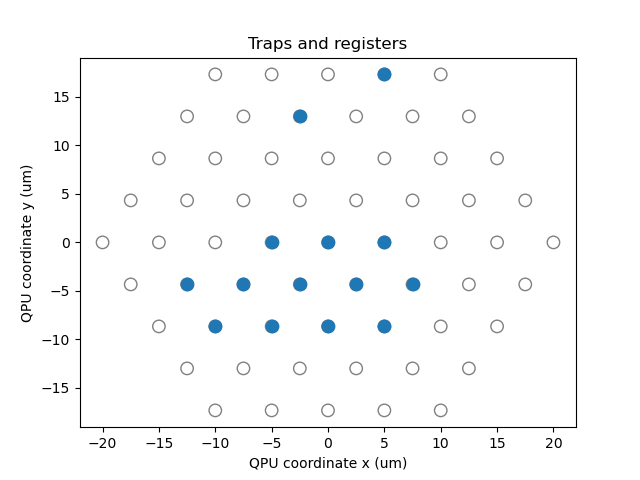} \\ \includegraphics[width=.4\linewidth,valign=t]{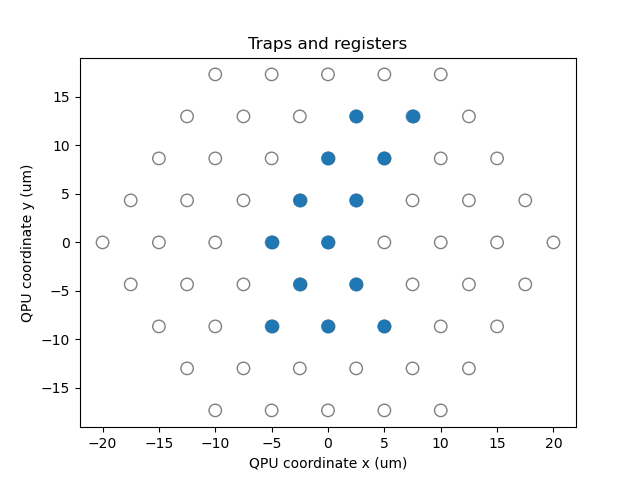} & \includegraphics[width=.4\linewidth,valign=t]{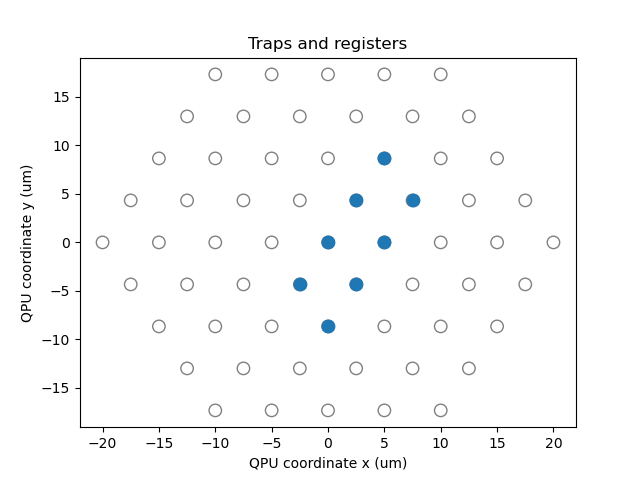}\\
\end{tabular}
\caption{Registers of Fig.~\ref{fig:local_algo_density_qubits}, reshaped to fit in the trap layout available on the QPU (see Fig.~\ref{fig:trap_layout}). The units correspond to the physical dimension of the registers.}
    \label{fig:traps_and_registers}
\end{figure}

Despite the algorithm being well-defined in any dimension, current generation neutral atom quantum computers can only operate on two-dimensional qubit registers. For this reason we cut the 3D-RISM density (which, by construction, is defined in three dimensions) into two-dimensional slices, and we apply the algorithm separately to each slice. For simplicity, we restrict ourselves to a small region of the protein around the two crystal water molecules. The plane corresponding to the first slice is defined by the axis connecting the two crystallographic water molecules and a third random direction. A series of slices is then produced from this first one in the direction normal to this plane, separated by a spacing of 0.5 \AA, for a total of 6 different slices. To smooth the 3D-RISM density in each slice, we apply a Laplacian of Gaussian (LoG) filter as provided in the SciPy Python library~\cite{scipy}, with a $\sigma = 8$ (\textit{a.u.}). The high-density regions of each smoothed-out slice are then covered with a qubit register. The process is depicted schematically in Fig.~\ref{fig:3DRISM_slicing}. Overall, this test implementation requires 4 to 14 qubits, which remains in the range of qubit count that can be classically emulated with high precision by a state vector solver with a realistic noise model. The density slices with associated qubit registers are shown in Fig.~\ref{fig:local_algo_density_qubits}. In order to mimic a physical implementation, we fit the registers in the same trap layout of Fig.~\ref{fig:trap_layout}, as shown in Fig.~\ref{fig:traps_and_registers}.

For each register, we design an adiabatic pulse to find the low energy states of the Rydberg Hamiltonian \eqref{eq:local_rydberg_hamiltonian}. The detunings $\Delta_i$ are computed from the $\Gamma_i$ coefficients of each problem Hamiltonian by using the following heuristic mapping:
\begin{equation}\label{eq:detuning_mapping}
    \tilde{\Delta}_i = \Gamma_i - \frac{1}{|N(i)|}\sum_{j \in N(i)} \Gamma_j, \quad \quad \Delta_i = \tilde{\Delta}_i \frac{\Delta_\text{max}}{\max_i |\tilde{\Delta}_i|}
\end{equation}
where $N(i)$ denotes nodes situated within a Rydberg blockade distance from node $i$, and $\Delta_{\max}$ is a fixed value. The mapping \eqref{eq:detuning_mapping} can be understood as a two-step process. First, each $\Gamma_i$ is shifted so that it is centered around the mean value of neighbouring nodes, obtaining the intermediate detunings $\tilde{\Delta}_i$. Then, the $\tilde{\Delta}_i$ are rescaled so that the maximum one corresponds to a certain fixed value $\Delta_{\max}$, obtaining the final value $\Delta_i$.

\begin{figure}[!htbp]
    \centering
\begin{tabular}{ll}
    \includegraphics[width=.4\linewidth,valign=t]{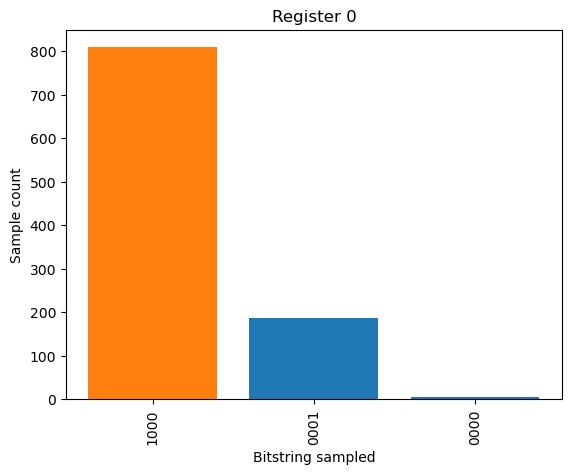} & \includegraphics[width=.4\linewidth,valign=t]{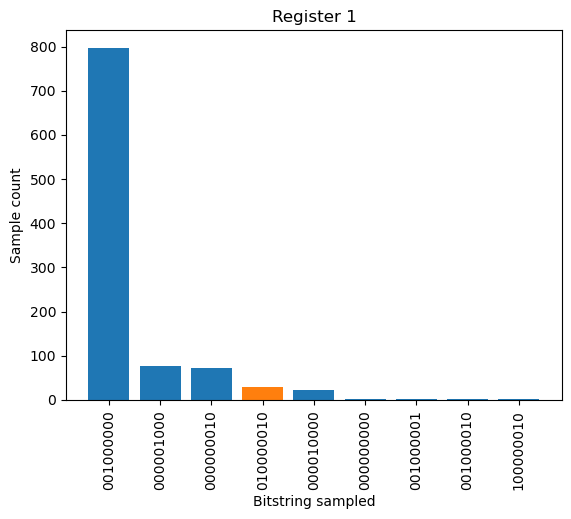} \\ \includegraphics[width=.4\linewidth,valign=t]{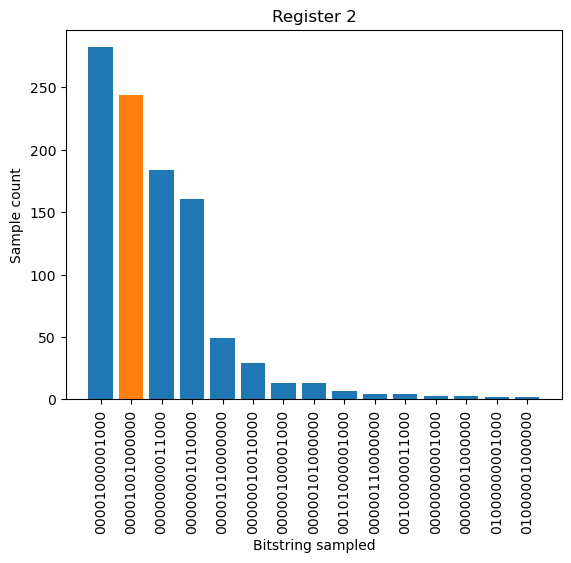} &
    \includegraphics[width=.4\linewidth,valign=t]{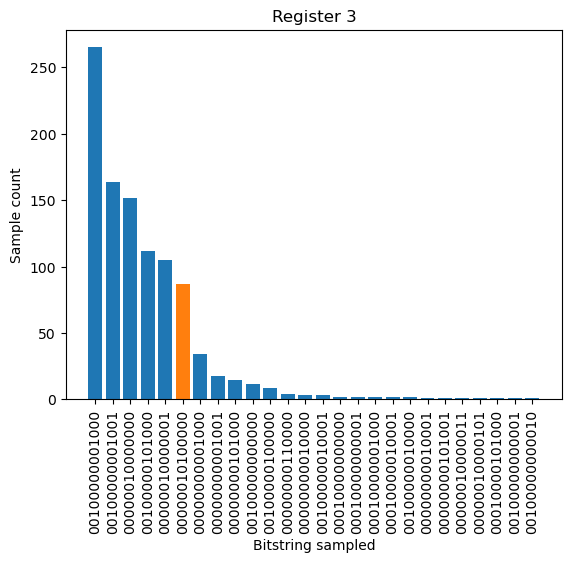} \\ \includegraphics[width=.4\linewidth,valign=t]{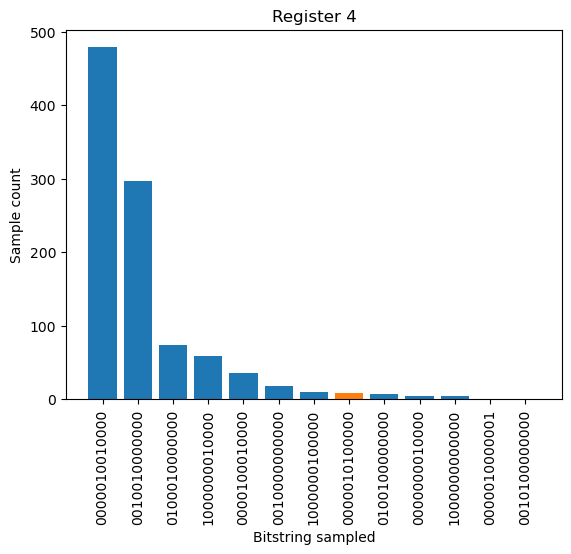} & \includegraphics[width=.4\linewidth,valign=t]{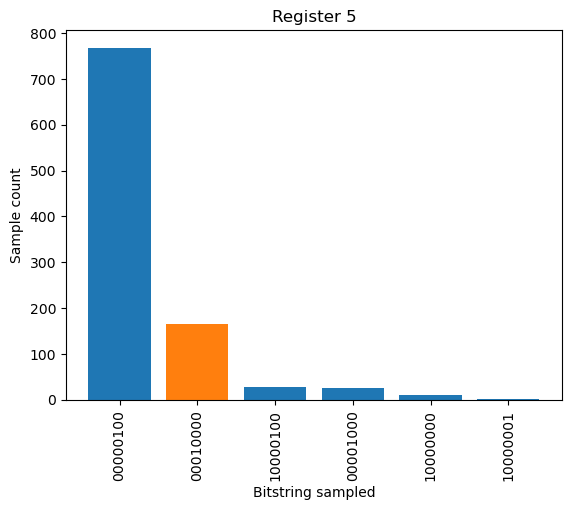}\\
\end{tabular}
\caption{Histogram of 1000 samples of the quantum state obtained with the local algorithm based on adiabatic evolution. Each plot corresponds to one of the registers shown in Fig.~\ref{fig:local_algo_density_qubits}. The orange bar corresponds to the bitstring with the minimal cost among the ones that were sampled, and it is taken as the solution of the water placement problem for that register.}
    \label{fig:local_algo_results}
\end{figure}

We simulate the resulting adiabatic dynamics with Pulser, and sample the final state 1000 times. For each computational basis state sampled, we calculate its cost according to the problem Hamiltonian, and select as a solution the one with the minimal cost. The resulting sampling, where the basis states are represented as bitstrings, is shown in Fig.~\ref{fig:local_algo_results} for all 6 slices. The orange bar in each plot identifies the bitstring corresponding to the best solution for that slice. The fact the bitstring with the lowest cost is not always the most sampled one is a consequence of the differences between the low energy spectrum of problem and Rydberg Hamiltonians. 

\textbf{Experimental measurement on QPU exploiting the local algorithm results.} The emulations presented in the previous section correspond to the capabilities of a neutral atom QPU capable of local addressing, which will be available in the next generation of neutral atom devices. An earlier prototype, with limited functionality, of such machines was tested here to confirm that solutions to the water placement problem belong to the space of physically allowed configurations of the machine.

In this simplified setting we use the same 3D-RISM densities and qubit registers presented in Fig.~\ref{fig:traps_and_registers} and Fig.~\ref{fig:traps_and_registers}. Each register was associated to the solution found by simulating the local algorithm on a classical computer, as reported in Fig.~\ref{fig:local_algo_results}. The qubit register is then reproduced in the QPU, and its quantum state is measured after evolving under a constant global laser pulse. Tunable parameters of the driving laser are its Rabi frequency $\Omega$, detuning $\Delta$ and duration. If $b_i$ indicates the bitstring encoding the solution to the water placement problem for the $i$-th register, and $\ket{b_i}$ is the associated computational basis state, we evaluate the expectation value of its projector $\ket{b_i}\bra{b_i}$ on the final quantum state, in a two-dimensional parameter space spanned by detuning and duration of the driving laser, while the Rabi frequency is fixed to a value that would enforce a Rydberg blockade between nearest-neighbours.

\begin{figure}[!htbp]
    \centering
    \begin{tabular}{@{}c@{}}
        \includegraphics[width=.7\textwidth]{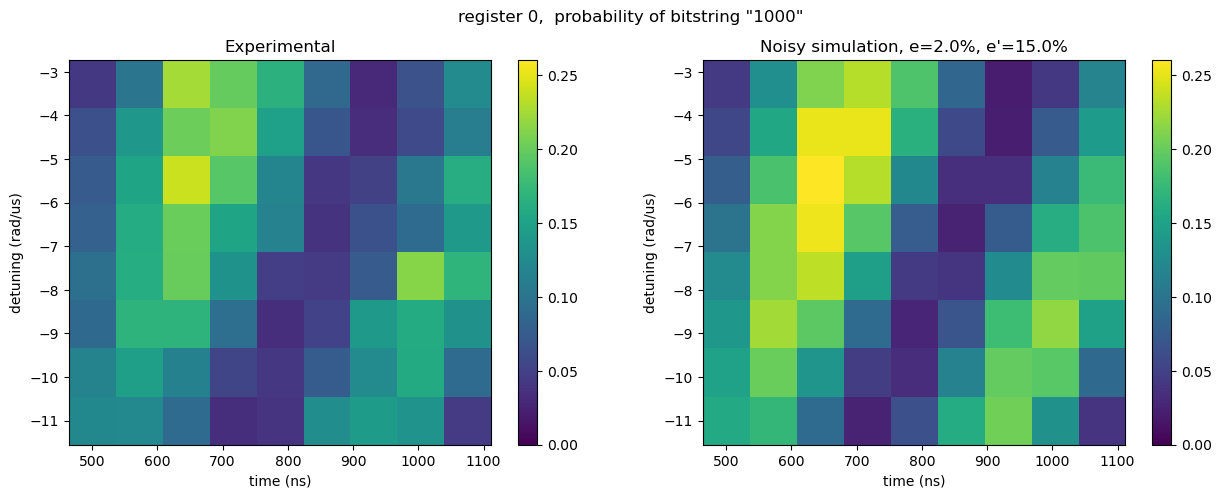} 
    \end{tabular}

\vspace{\floatsep}

    \begin{tabular}{@{}c@{}}
        \includegraphics[width=.7\textwidth]{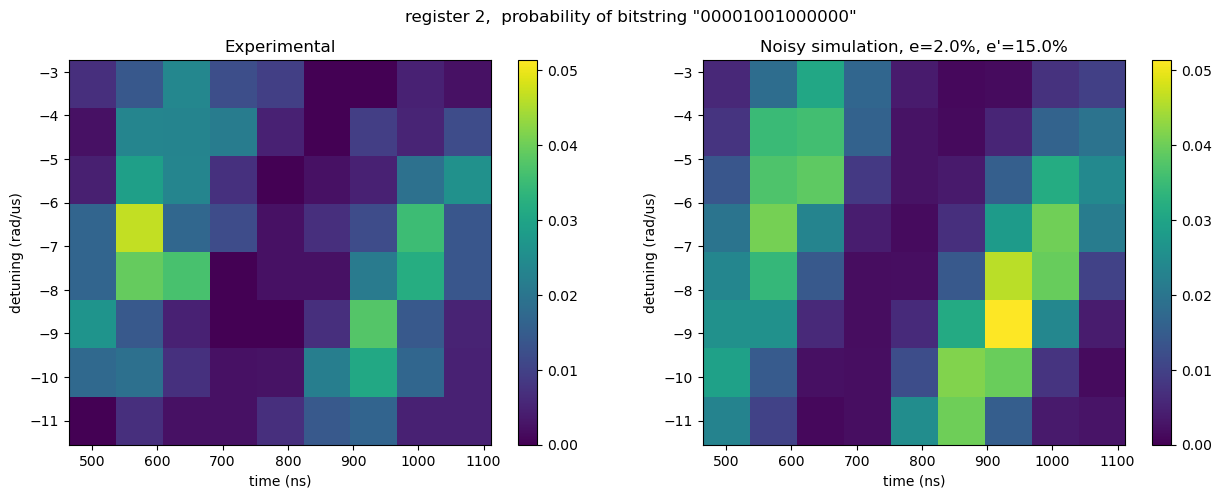} 
    \end{tabular}

\vspace{\floatsep}

    \begin{tabular}{@{}c@{}}
        \includegraphics[width=.7\textwidth]{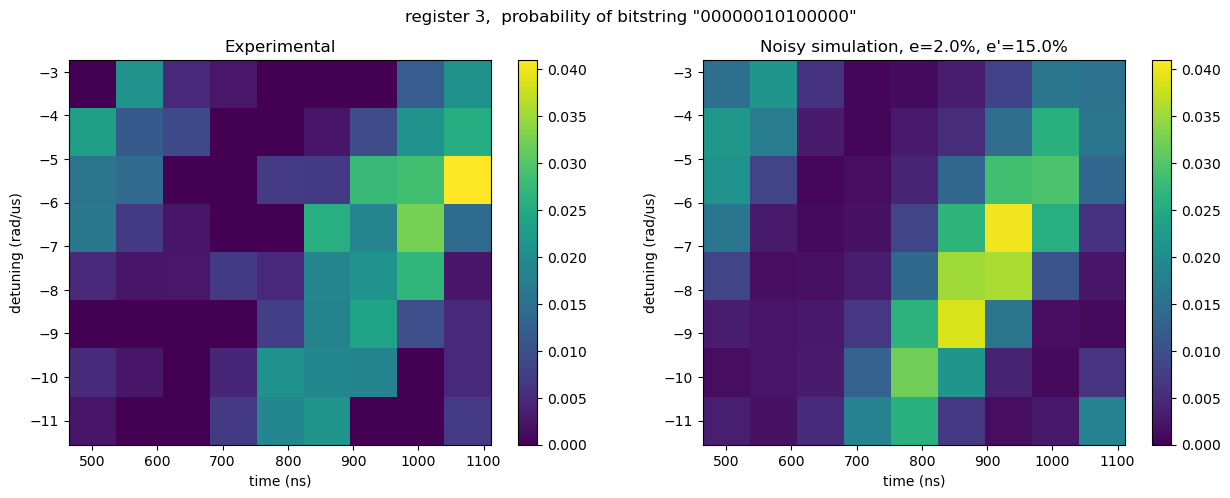} 
    \end{tabular}

\vspace{\floatsep}

    \begin{tabular}{@{}c@{}}
        \includegraphics[width=.7\textwidth]{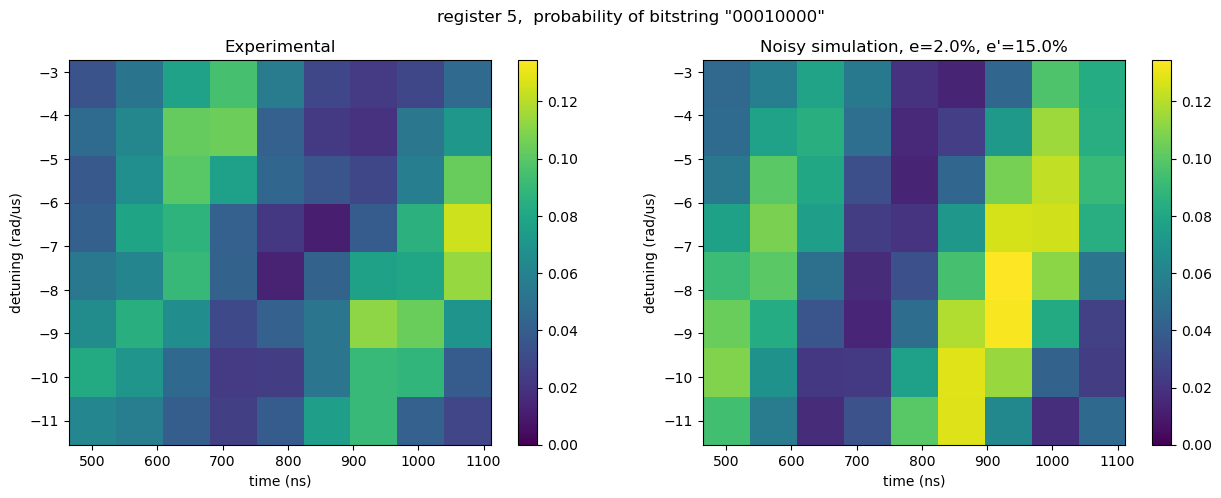} 
    \end{tabular}

    \caption{Expectation value of the projector operator $\ket{b_i}\bra{b_i}$ associated to the solutions $b_i$, $i=1,2,3,4$ of the four of the  3D-RSM 2D slices . The area scanned is a two-dimensional region parametrized by pulse duration (horizontal axis) and detuning (vertical axis) of the driving laser. The plot on the left represents the experimental values, while the plot on the right represents the values obtained from a classical simulation of the Rydberg Hamiltonian with measurement errors $\epsilon=0.02$ and $\epsilon'=0.15$ related to the probability of false positive and false negative detection.}
    \label{fig:best_landscape}
\end{figure}

The experimental measurements of the projector for all four registers are shown in Fig.~\ref{fig:best_landscape}, together with theoretical expectations given a reasonable estimate for false positive and false negative error rates on the machine. Each point in the 2D plane is calculated as an average over roughly 500 measurements. On a qualitative level, the experiments can be seen to be compatible with theoretical predictions obtained from emulation. These results indicate that meaningful solutions to the water placement problem belong to the space of physically allowed configurations of the QPU. To our knowledge, such use of Rydberg physics had not been explored before in a real quantum physics experiment. Furthermore, the experiment shows that probability landscapes in this simplified setting can be accurately resolved by the machine, even in the absence of error correction or error mitigation techniques. This is a key result knowing the small magnitude of the involved probabilities that are capped at only a few percent in certain systems.

\begin{figure}
    \centering
    \includegraphics[width=0.7\textwidth]{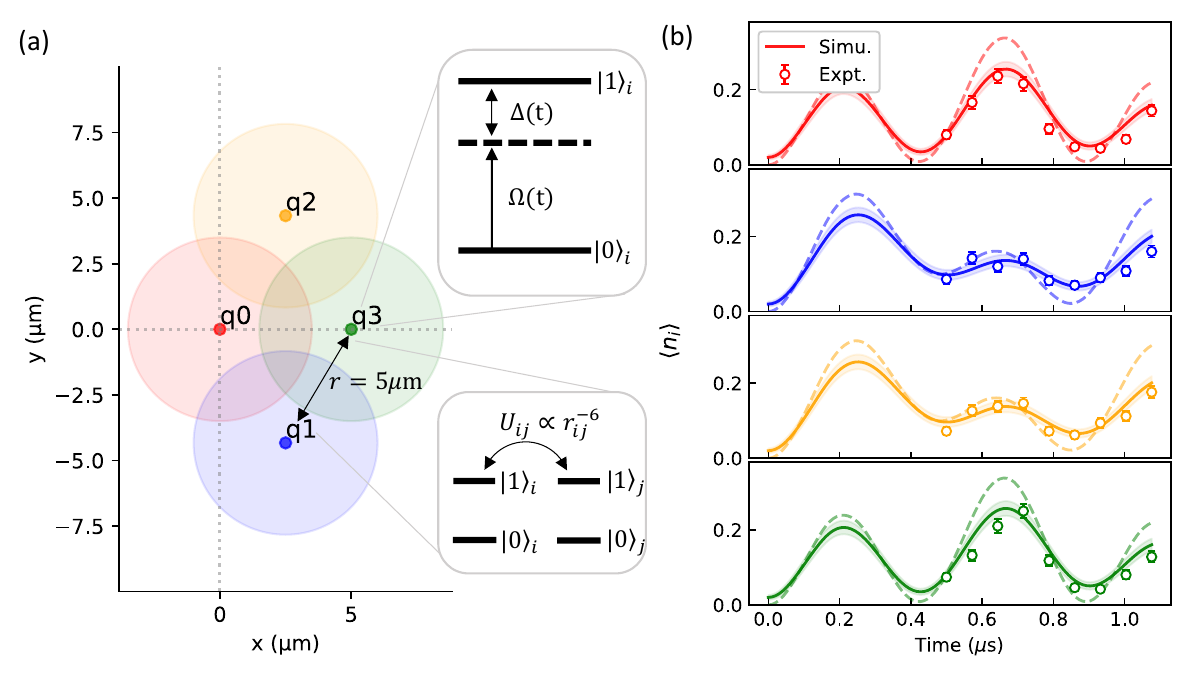}
    \caption{(a) A laser pulse of constant $\Omega$ and $\Delta$ is applied for various times on the first atomic register comprising of 4 qubits (colored). (b) For a fixed detuning $\Delta/2\pi=-0.87$ MHz, the expectation value of the occupation of each qubit site, $\langle n_i\rangle$, is plotted for noiseless simulation (dashed), noisy simulation (line) and experimental data (dot). The error bars and uncertainty regions are the standard deviation computed over $N_{shots}=500$ samples, $\sigma_{n_i}=\sqrt{\langle n_i\rangle(1-\langle n_i\rangle)/N_{shots}}$.}
    \label{fig:fit_occupation}
\end{figure}

In order to better understand and quantitatively validate the experimental data produced by the set-up, we considered a more sophisticated error model. Calibration of the control devices can only be achieved with finite precision, resulting in static uncertainties in global spacing of the atomic array ($\approx 1\%$) or in the spatial homogeneity of $\Omega$ on this array ($\approx 4\%$). In addition, fluctuations of laser intensity induces a shot-to-shot variation of $\Omega$ ($\approx 5\%$). While the laser frequency can be set with high precision, variation in $\Omega$ indirectly alters $\Delta$, resulting in small detuning shifts of the order of $2\pi\times 0.06$ MHz. Decay processes are also taken into account by solving the Master equation with an effective decay rate $\Gamma_{eff}/2\pi$. Finally, the measurement phase is inherently flawed by
several physical processes like atomic losses due to background-gas collisions or Rydberg state finite lifetime, whose effects can all be encompassed as first approximation into two detection error terms, $\varepsilon$ and $\varepsilon^\prime$. The various values of experimental and noise parameters are usually fitted by comparing the expectation value of easy-to-access observables between simulated and experimentally acquired data. For instance, emulating the dynamics described by the Hamiltonian \eqref{eq:global_hamiltonian} in presence or absence of noise processes and measuring the occupation $\hat{n}_i$ at each site enables to estimate for the magnitude of the various error sources. An example of fitting curves at a given detuning $\Delta$ for the first register is shown in Fig.~\ref{fig:fit_occupation}. While the errors on calibration parameters are directly measured on the experiment, the remaining effective noise parameter are found to be $\Gamma_{eff}/2\pi=0.05$ MHz, $\varepsilon=2\%$ and $\varepsilon^\prime=18\%$.

Overall, given the experimental errors sources that were discussed, the experimental data appear fully compatible with the emulated ones (see Fig.~\ref{fig:best_landscape} and ~\ref{fig:fit_occupation}) confirming the viability of the proposed approach on an actual cold atom device.
We want to stress here that thanks to the quantum nature of the algorithm, two water molecules are guaranteed to never be placed too close to each other, a constraint that is imposed by hand in classical approaches such as Placevent~\cite{Placevent} or GAsol~\cite{fusani_optimal_2018}  with the risk of incurring in suboptimal local solutions or non-ergodicity. This constitutes one of the main advantage of the algorithm.

\textbf{Beyond a Local Algorithm: Variational Algorithm using a global Rydberg Hamiltonian.} As we discussed, technical limitations of the present hardware prevent us from using a local Hamiltonian as in equation~\eqref{eq:local_rydberg_hamiltonian}. However, through emulation, we can still plan for an algorithm that will be able to run on more short-term devices using a global Rydberg Hamiltonian:
\begin{equation}\label{eq:global_hamiltonian}
    \hat{H}(t) = \Omega(t)\sum_{i=1}^M\hat{\sigma}^x_i - \Delta(t)\sum_{i=1}^M \hat{n}_i +\sum_{i<j=1}^{M} \frac{C_6}{r_{ij}^6} \hat{n}_i \hat{n}_j.
\end{equation}

Such formulation allows to  further approximate the problem Hamiltonian in eq.~\eqref{eq:problem_hamiltonian}, since the linear term in the qubits excitation is no longer local. A variational procedure is then established using, as cost function to be minimized the problem Hamiltonian in eq.~\eqref{eq:problem_hamiltonian}, so to solve

\begin{equation}\label{eq:algorithm-problem}
    (\Omega(t),\Delta(t))^{*}= \displaystyle\argmin_{\Omega(t),\Delta(t)} \bra{\Psi^{\Omega,\Delta}} \hat{I}^2 \ket{\Psi^{\Omega,\Delta}}.
\end{equation}
This algorithm is therefore part of the Variational Quantum Algorithm (VQA) family
The cost function $\bra{\Psi^{\Omega,\Delta}} \hat{I}^2 \ket{\Psi^{\Omega,\Delta}}$ is evaluated summing up the contributions of each basis state sampled from a trial wave function, obtained from the laser parameters $(\Omega(t),\Delta(t))$ produced during each optimization step. The full VQA developed for this scope is described in Algorithm~\ref{alg:HQC_IsG}.

\begin{algorithm}[htpb]
\caption{Hybrid quantum-classical Variational Quantum Algorithm (VQA) using global lasers, with cost function issued from the Ising model, namely $\bra{\Psi^{\Omega,\Delta}} \hat{I}^2 \ket{\Psi^{\Omega,\Delta}}$.\newline
\textbf{inputs}: reference 2D density $g(\vec{r}) : \mathcal{C} \rightarrow \mathbb{R}$ and set of qubits position $Q := \{\vec{q}_i\} \subset \mathcal{C}$, with $\mathcal{C} \subset \mathbb{R}^{2}$. $n_c$ and $n_r$ are parameters for the Bayesian optimization procedure, being, respectively, the total number of cycles performed to generate a single state $\ket{\psi}_{k=1,\dots,n_c}$ and $n_r<n_c$ is the number of cycles initiated with randomized values of laser parameters $\Omega$ and $\Delta$.\newline
\textbf{outputs}: water molecule positions $\mathcal{W}$ and their number $\mathcal{N}$.
}
\label{alg:HQC_IsG}
\begin{algorithmic}[1]
\Statex
\Procedure{VQAgo}{$g(\vec{r}), \{\vec{q}_i\}, n_r, n_c$}
    \Statex
     \State $k \gets 0$
    \While{$k<n_c$}
        \If{$k \leq n_r$}\Comment{The first $n_r$ iterations are uniformly randomized}
         \State  $\Omega_k$, $\Delta_k \gets$ \Call{Random}{$\mathcal{U}$} 
        \Else
           \State $\Omega_k, \Delta_k \gets$ \Call{BayesianMinimization}{$\{\Omega_l, \Delta_l\}_{0\le l < k}$, $\bra{\Psi^{\Omega,\Delta}} \hat{I}^2 \ket{\Psi^{\Omega,\Delta}}$}
        \EndIf
        \State $\ket{\Psi_k} \gets $ \Call{QuantumEvolution}{$\{\vec{q}_i\}, \Omega_k, \Delta_k$}
        \State $J_k \gets \bra{\Psi_k} \hat{I}^2 \ket{\Psi_k}$
        \Comment{Collection of $J_k \forall k$ to evaluate later best laser parameters}
    \EndWhile
        \State $\Omega^*, \Delta^* \gets$ \Call{ArgMin}{$\{J_k\}$} \Comment{Select the $\Omega(t)$ and $\Delta(t)$ that minimize the cost function according to eq.~\eqref{eq:algorithm-problem}}
        \State $\ket{\Psi^*} \gets $ \Call{QuantumEvolution}{$\{\vec{q}_i\}, \Omega^*, \Delta^*$}
        \State $\ket{e^*} \gets$ \Call{BestBasisState}{$\ket{\Psi^*}$} \Comment{Basis state whose GMA best represents $g(\vec{r})$ sampled from the optimized $N$-qubits wavefunction}
    \Statex
    \State $(\mathcal{W}, \mathcal{N}) \gets$ \Call{PositionsFromState}{$\ket{e^*}$}
    \Comment{$(\mathcal{W}, \mathcal{N})$ are defined in eq.~\eqref{eq:wat_numcoord}}
    \State \textbf{return} $(\mathcal{W}, \mathcal{N})$ \Comment{Best placement of water molecules in, \textit{e.g.}, target protein cavity}
    \Statex
\EndProcedure
\end{algorithmic}
\end{algorithm}

$\ket{\Psi^{\Omega,\Delta}}$ in Eq.~\eqref{eq:algorithm-problem} is a short notation for a quantum state obtained from the evolution dictated by the Hamiltonian in eq.~\eqref{eq:global_hamiltonian} for a specific set of laser parameter $(\Omega(t),\Delta(t))$.

The objective of the numerical procedure is to maximize the probability of sampling the basis state~\eqref{eq:bs_best}, from the optimized $N$-qubits wavefunction. This corresponds to finding one or more configurations of excited qubits best representing the 3D-RISM density distribution as a sum of gaussian distributions. The final output is still described by the quantities in Eq.~\eqref{eq:wat_numcoord}.

To test the performances of the algorithm we use again simple synthetic densities to limit the number of qubits to employ, so that a cycle of optimization, emulating multiple times the quantum evolution of the system on the CPU, is performed for each test case.
The results are reported in SI (see section~\ref{num-sim-support}). They show that the algorithm is able to give the correct positions of the Gaussian distributions centroids. Due to the high computational cost involved in the emulation of a numerical solution of the time-dependent Shr{\"o}dinger equation, we could not extend our tests towards larger systems.

\begin{figure}[!htbp]
\begin{center}
\includegraphics[width=0.59\textwidth]{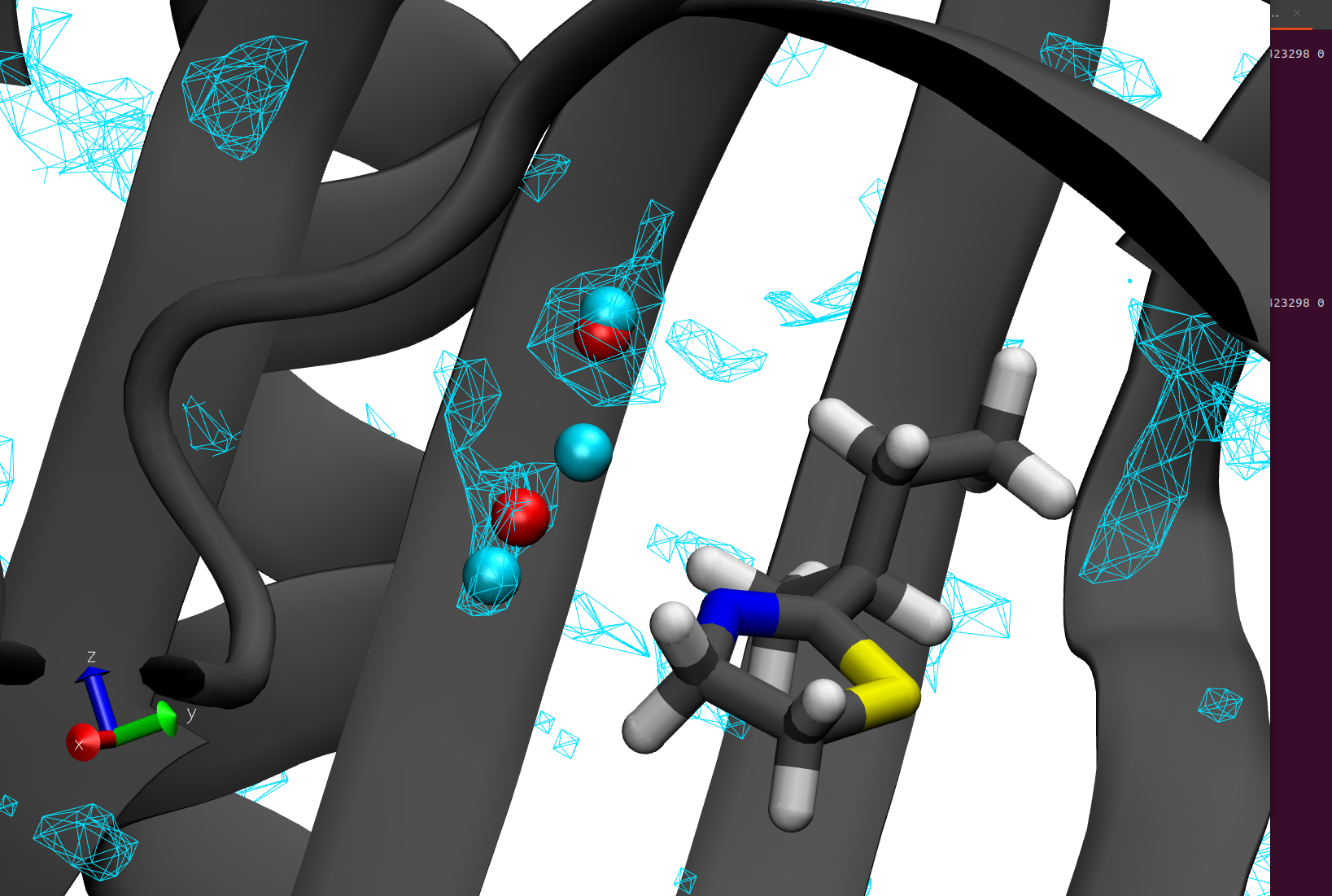}
\end{center}
    \caption{MUP-I protein complexed with the 2-sec-butyl-4,5-dihydrothiazole, inside view of the protein binding pocket. Red atoms are oxygen atoms of the crystal water molecules, and the cyan one are the three water molecules placed by the HQC algorithm . The 3D-RISM density isosurface (isovalue 5.5) is represented in wireframe style, in cyan. }
    \label{fig:MUP_T1res}
\end{figure}

Furthermore, We used as an additional test case one of the slices obtained from the 3D-RISM density, and already presented for the QAE approach. The sliced density, together with the qubit register, are shown in Fig.~\ref{fig:MUP_strct} (panel c). We perform 50 cycles of Bayesian optimization, using 200 samples to represent each wavefunction produced by the corresponding set of laser parameters. The gaussians used to dress the excited qubits, so to assign scores to the corresponding bitstring, have amplitudes proportional to the local value of the 3D-RISM density and an uniform variance $\sigma^{2} = 5$ (\textit{a.u.}).
From the resulting final wavefunction, we obtain the 3D coordinates of the oxygen atoms of the placed water molecules. In Fig.~\ref{fig:MUP_T1res} we report the best water molecules configuration found by the algorithm (cyan atoms), corresponding to the best bitstring $\ket{101001}$ found in the optimization. Also two excited qubits states, as $\ket{010100}$ and $\ket{010001}$ show good sampling probability (see corresponding histogram).

Additional emulated experiments are performed on all the other slices and the results are reported in the SI. The algorithm exhibits equivalent performances in all these cases. Again, we stress that this variational quantum algorithm, like the version using the local lasers, prevents two water molecules from being placed too close to each other, a constraint that is imposed by hand in classical approaches.

\textbf{Conclusion and perspectives.} In this work we presented two new quantum algorithms, able to sample equilibrium solvent configurations within proteins. 
We proposed a first fully local quantum adiabatic evolution version to be used on next generation devices whereas, a second version, belonging to the variational quantum algorithm family of algorithms has also been introduced as a viable short-term alternative.  
This new class of algorithms correspond to quantum versions of the 3D Reference Interaction Site Model (Quantum-3D-RISM or Q-3D-RISM) since they use continuous solvent distributions and are able to efficiently locate density distributions' maxima using a neutral atom quantum computer. As a proof of concept, both algorithms have been shown to successfully be able to locate density maxima in non-trivial densities. In the case of the second VQA algorithm, a classical optimization is performed to find the best set of laser pulses, capable of producing the expected distribution of maxima. A more simplified version was implemented on a real neutral atom QPU, which represents to the best of our knowledge the first application of an analog quantum algorithm to the protein solvation problem.
Presently, we limited ourselves to a qubit count of 14. This number was constrained by the present machine layout (see Fig.~\ref{fig:trap_layout} and Fig.~\ref{fig:traps_and_registers} ), by the state vector emulation capabilities (roughly 16 qubits can be emulated easily with Pulser) and the need of coupling Pulser and the actual QPU for the quantum adiabatic evolution model implementation. In that connection, concerning the local algorithm, future availability of time-dependent pulses and local addressing will totally remove the need of the Pulser emulation to prepare QPU data. In the same line, only time-dependent pulses are required for the VQA version of Quantum-3D-RISM: it should enable us to perform larger simulations at short-term since the upgrade of the PASQAL's machine is ongoing. Moreover, since the quantum versions of 3D-RISM comes with the native advantage of preventing water molecules from being placed on top of each other, it will be  interesting to compare the performances of the classical and quantum versions of 3D-RISM since the next QPU implementation will provide us with the possibility to test at large scale the accuracy of such techniques. To do so, a careful study of the machine noise will be necessary but one key advantage of analog computing is to exhibit relatively constant noise levels with increasing system size making us optimistic about the prospect of QPU simulations encompassing a high number of qubits. At that stage, it will be then possible to prepare any molecular system with such algorithms in order to couple them to state-of-the-art molecular dynamics engines\cite{THP,THP-GPU} for further properties evaluations.  Overall, this Quantum-3D-RISM (Q-3D-RISM) family of algorithms demonstrates promises in predicting the solvation structure within biomolecular systems of interest for drug discovery applications, providing concrete use cases for the application of analog quantum computing in life sciences.

%
%
%

%
%
%
%

%
%
%
%

\section*{Appendix A: Quantum computational resources}
For the quantum computing sequences, we make use of Fresnel, an industrial neutral atom QPU made of single 87Rb atoms trapped in arrays of optical tweezers, conceived and manufactured by PASQAL. We operate the QPU in the Ground-Rydberg qubit basis with global analog control~\cite{henriet2020quantum}. The qubits are encoded into the ground state $\ket{0} = \ket{5S_{1/2}, F = 2, m_F = 2}$ and a Rydberg state $\ket{1} = \ket{60S_{1/2}, m_J = 1/2}$ . This effective two-level system is addressed with a two-photon laser excitation through an intermediate state $6P_{3/2}$. The first (respectively second) photon excitation is generated by a 420-nm (1013-nm) laser beam. Details about the Pulser control software used to program the experiment can be found in~\cite{pulser}.

\section*{Acknowledgements}

This work was made possible thanks to the Pack Quantique grant from région Ile de France and GENCI, project ACQMED (convention N°20012758). Funding from the European Research Council (ERC) under the European Union's Horizon 2020 research and innovation program, project EMC2 (grant agreement N°810367), is also acknowledged (J.-P. P) as funding from PEPR Epiq (ANR-22-PETQ-0007) and HQI programs. 

\subsection*{Authors Contribution Statement}
M.D, D. L performed simulations and contributed new code;\\
M.D, D.L, N. G, P. M, S. A, J. S, J. F, L.-P. H, J.-P. P contributed new methodology (theory);\\ 
The Fresnel team, L. H designed and performed the experiments; \\
M.D, D.L, the Fresnel team, N. G., L.-P.H, L. H, J.-P. P analyzed data;\\ 
M.D, D.L, part of the Fresnel team, L.-P. H, L. H., J.-P. P wrote the paper with the input of all authors;\\
L. H and J--P. P designed the research.
\\ \\
The Fresnel team (Pasqal):
J. ARMOUGOM, D. BENVENUTTI, L. BEGUIN, L. BOURACHOT, J. BRIAND, C. BRIOSNE FREJAVILLE, N. CARREZ, T. CARTRY, A. CHARPENTIER, D. CLAVEAU, L. COLIN, G. COURNEZ, L. COUTURIER, J. DE HOND, S. DESIRE, A. DUMAS, S. DUTARTRE, P. FAVIER, G. FIRENZE, D. KACZOR , C. HAMOT, G. HERCE, J. HEURTEBIZE, V. HULLY, B. LABARRE, L. LASSABLIERE, H. LE BARS , L. LECLERC, A. LINDBERG, G. MERIAUX, F. NAMBI, T. PANSIOT, G. PARIENTE, J. PELLEGRINO, L. PONSOT, S. ROCHE, H. SILVERIO, G. VILLARET, J.-M. WIPFF.

\bibliography{bibliography.bib}

\end{document}


\flushbottom
\maketitle
\section*{Supplementary Information}
\section{Numerical simulation on CPUs for model systems with a QAE}
\label{sec:num-sim-supportQAE}

We briefly recall here that, given an array of $M$ quits, the result produced by either the VQA or the QAE algorithms is the position of $N$ molecules placed in correspondence to $N$ of the $M$ excited qubits. The results is coherent with the Rydberg blockade constraint and is the best at reproducing the reference density distribution. 

\begin{figure}[!htbp]
\begin{center}
\subfloat[]{\includegraphics[width=0.40\textwidth]{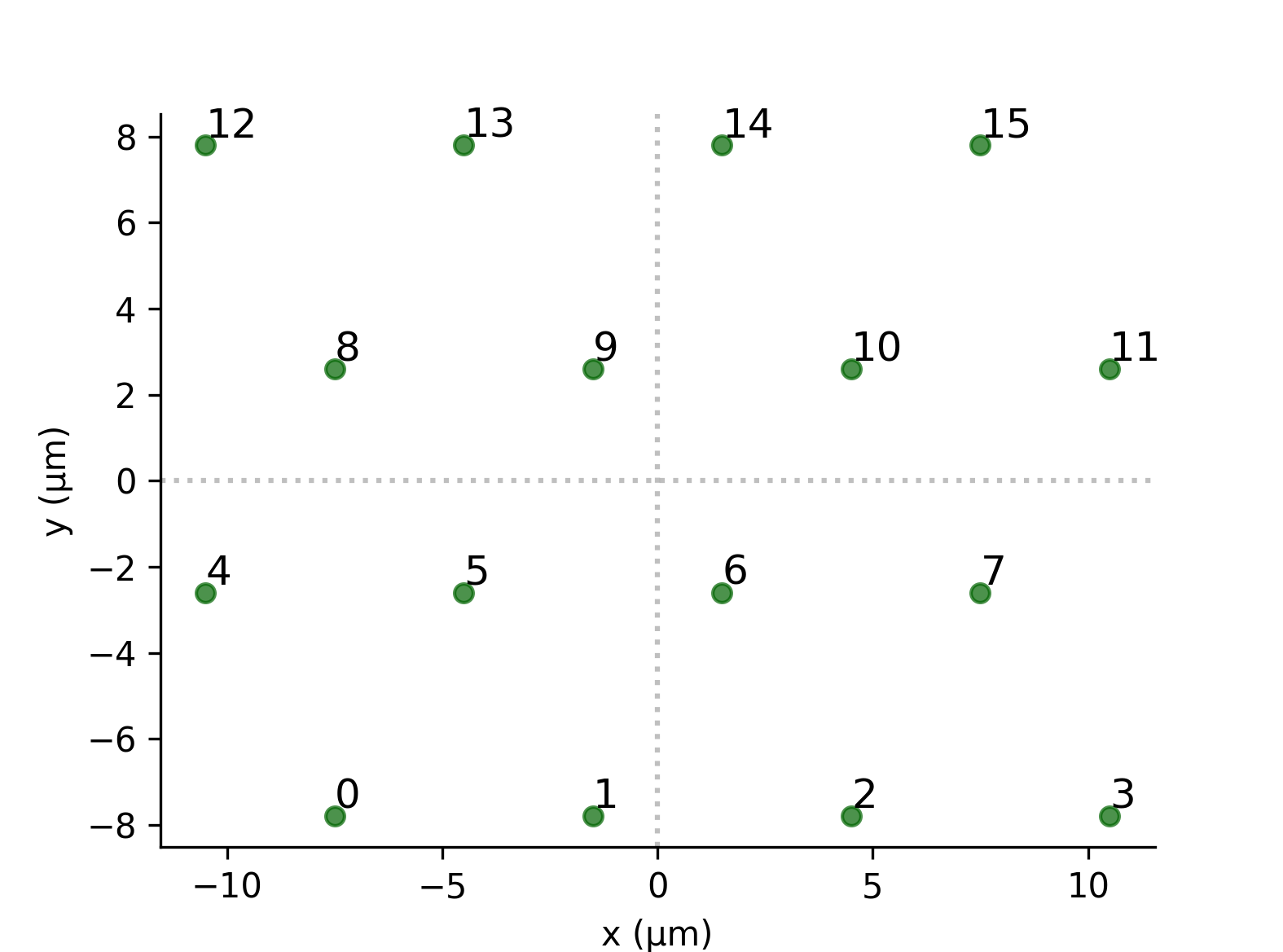}}
\subfloat[]{\includegraphics[width=0.59\textwidth]{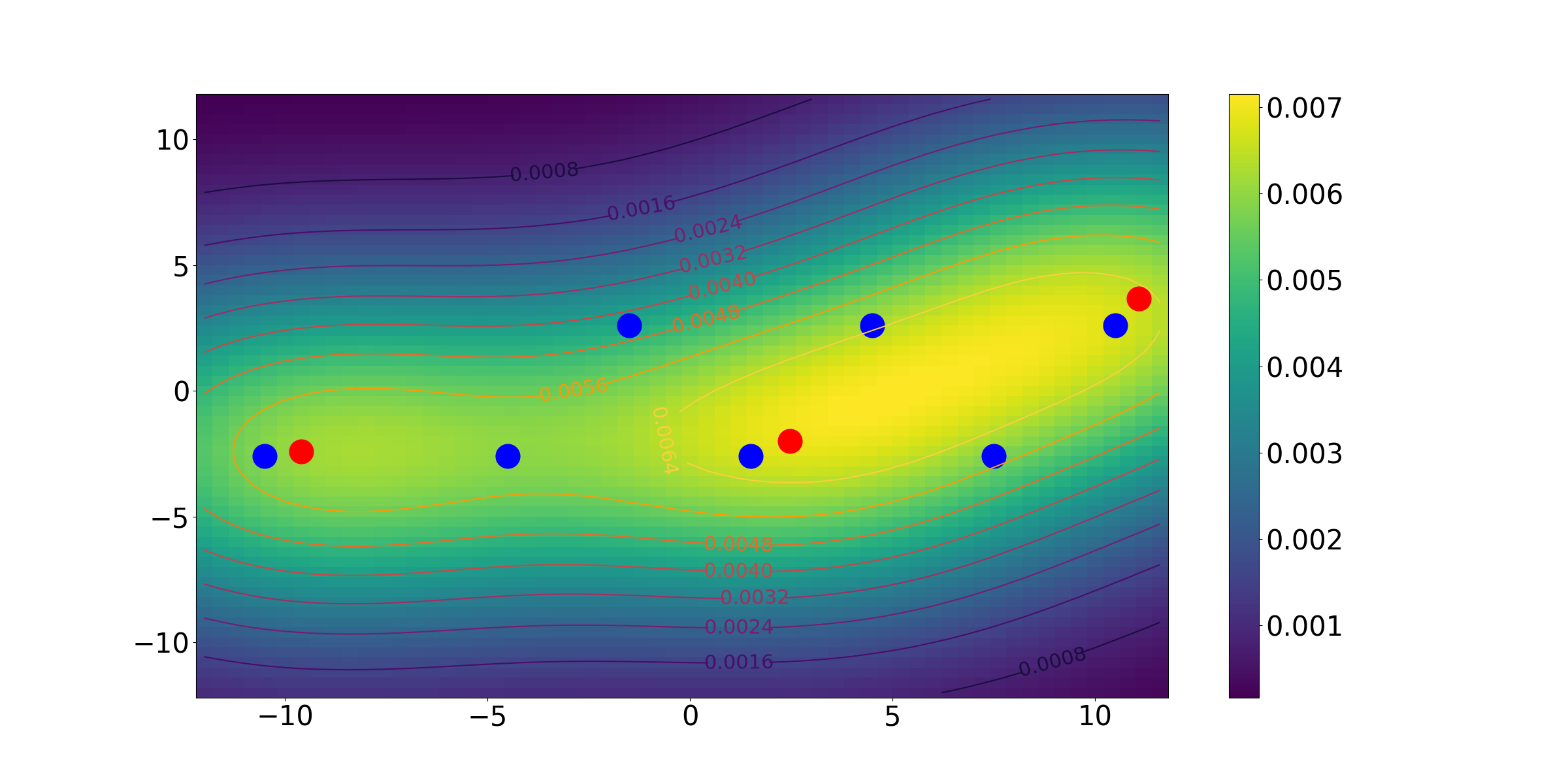}}
\end{center}
    \caption{(a) 2D qubits array used for the tests on the emulator; (b)  example of 2D density generated as a sum of gaussians; three gaussian distributions are summed up, and the respective centers are marked with red dots. Blue dots represents the only qubits, from the initial array in (a), which are placed in a region where the density is sufficiently high. The axes of the graph are spatial coordinates, in $\mu$m, while the colour palette from deep blue to yellow is used to mark the density level; level curves are also drawn on the same graph.}
    \label{fig:exampleD}
\end{figure}

The qubit array is initially prepared on a 2D plane, as if one was to prepare it for an experimental set-up (see panel a, Figure~\ref{fig:exampleD}). Ideally, the origin of the reference system is the focus of the laser that one would use in a real experiment. This initial qubit array is then reduced as shown in Figure~\ref{fig:exampleD}, (panel b) based on a local density threshold. The 2D density is represented on the same plane where the qubits have beed placed initially. This procedure is used to reduce the number of qubits to emulate, so to dramatically reduce the computational cost.

The quantum adiabatic evolution (QAE) algorithm cannot be implemented on the current generation of neutral atom QPUs because of the lack of local addressing. As stated already in the main text, in other words, instead of the local terms $\Omega_i(t)$ and $\Delta_i(t)$, one only has access to a global control field $\Omega(t)$ and $\Delta(t)$. On the other hand, the local algorithm can be tested using a classical emulator. 

For these and the other emulations performed on a classical CPU (\textit{vide infra}) we used an AMD EPYC 7742 64-Core Processor CPU (3368.100 MHz).

\begin{figure}[!htbp]
\begin{center}
\subfloat[]{\includegraphics[width=0.5\textwidth]{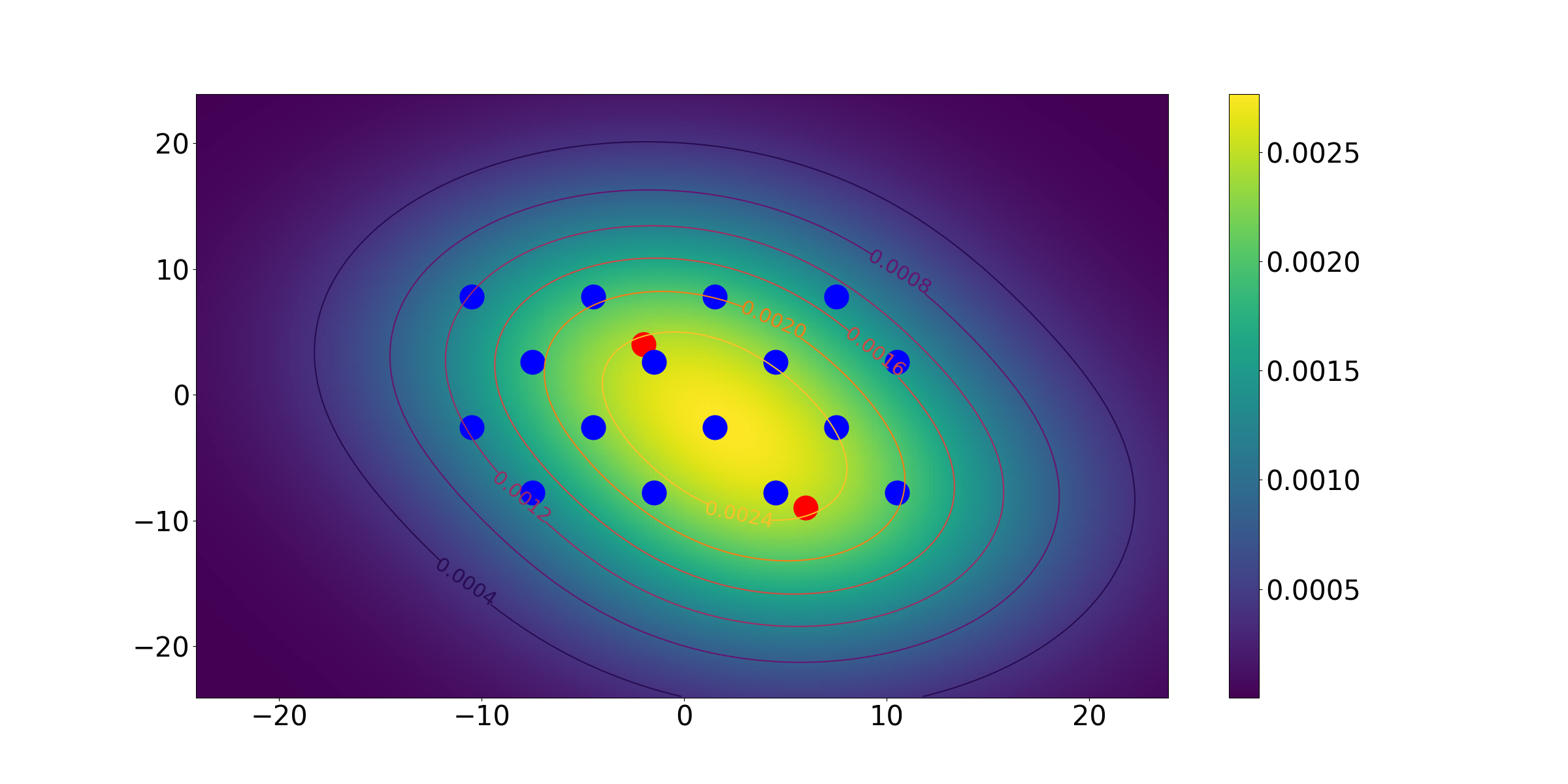}}
\subfloat[]{\includegraphics[width=0.5\textwidth]{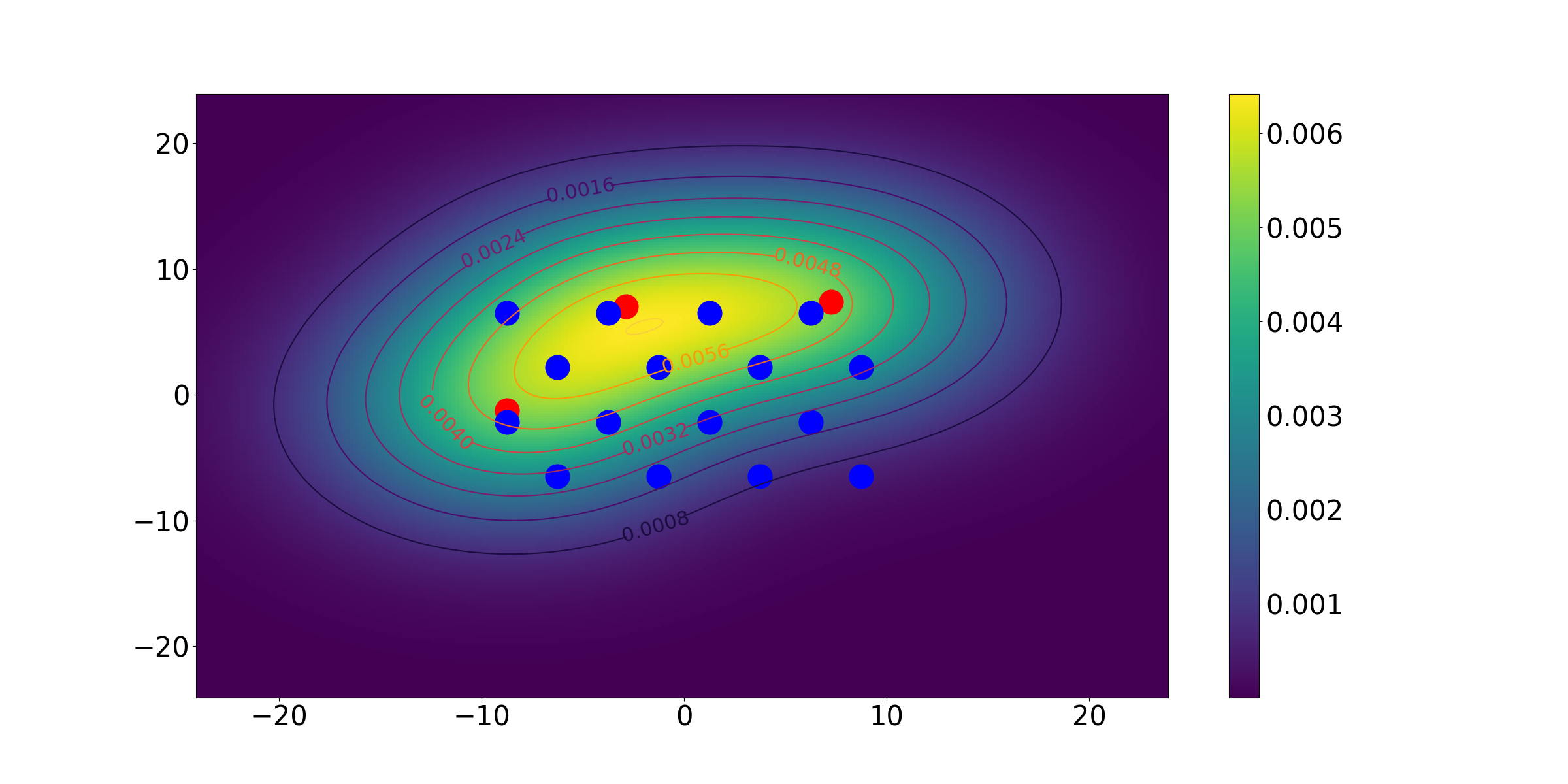}} \\
\subfloat[]{\includegraphics[width=0.5\textwidth]{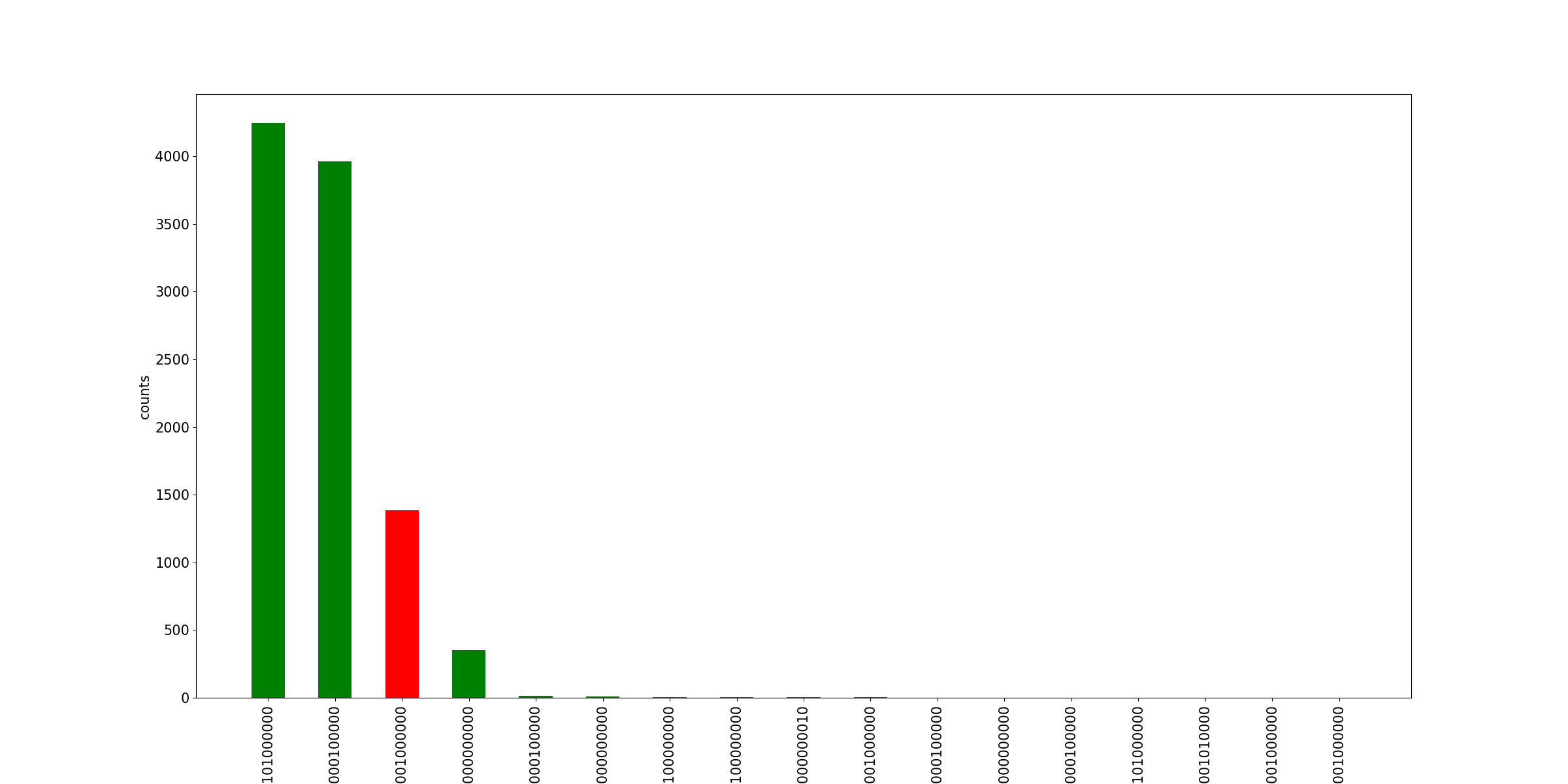}}
\subfloat[]{\includegraphics[width=0.5\textwidth]{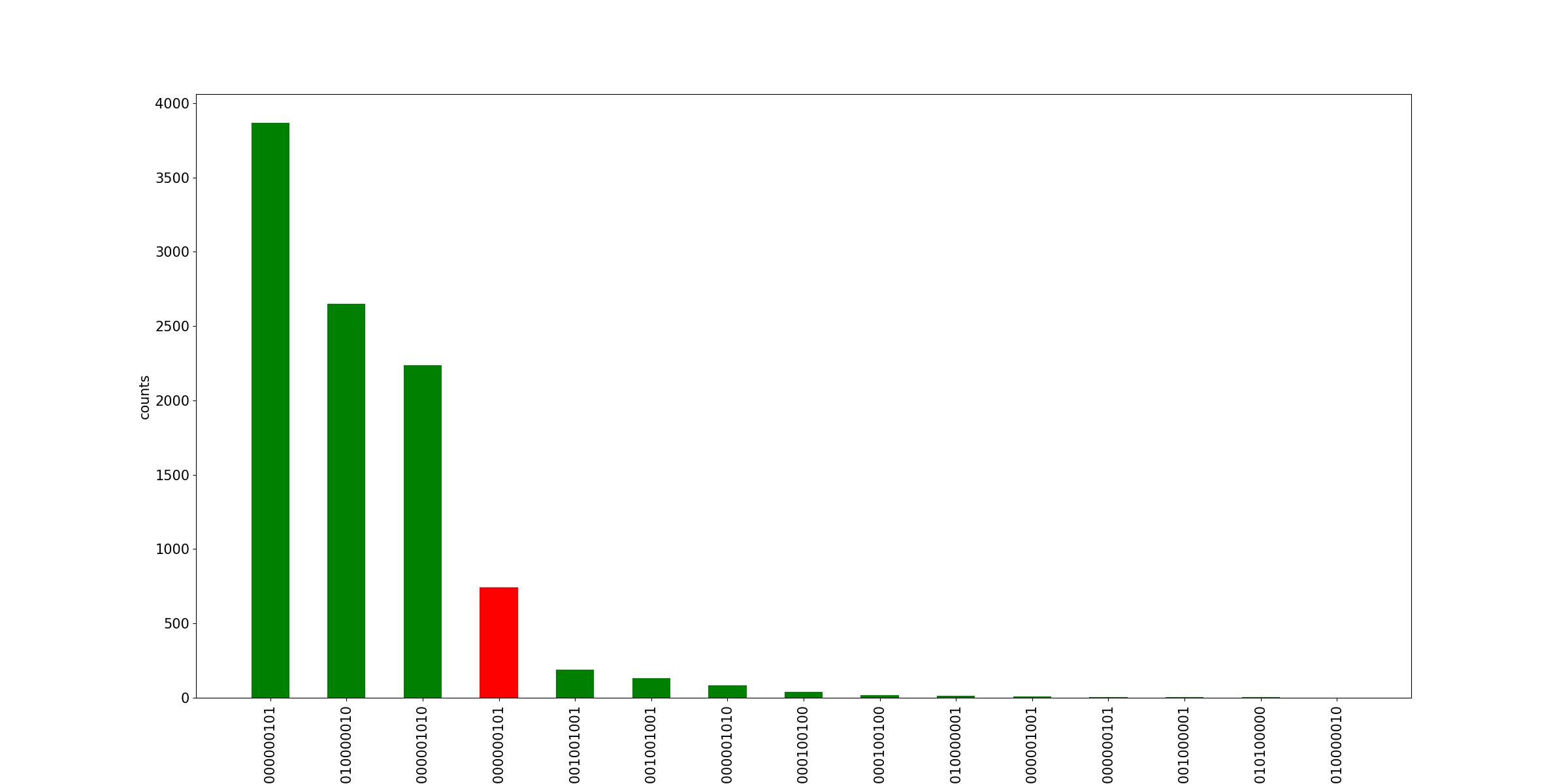}}
\end{center}
    \caption{Representation of the test synthetic sum of gaussian densities in 2D used to the the local laser based algorithm. The density maps, with level curves, are represented together with the qubits array (blue dots) built to address the location of the gaussian components. The centroid of each gaussian component is highlighted in red}
    \label{fig:QAE_2D_dens}
\end{figure}

To this aim we used the two 2D densities reported in Fig.~\ref{fig:QAE_2D_dens} (a and b subfigures).  We will refer in the following to those densities as \textit{synthetic densities}, since we generated them as a sum of gaussian distributions with known components. For each of case 16 qubits are used, placing them uniformely on the 2D densities.
The output of each emulation is essentially a sampling of the optimized $N$-qubits state $\ket{\Psi^{\{\Omega_{i}(t'),\Delta_{i}(t')\}}}$. This amount to sample the different qubits bitstrings, differing from total number of excitations and excited qubits, composing the state. The corresponding states obtained for the two test systems are reported in ~\ref{fig:QAE_2D_dens} c and d. The probability of those states best representing the gaussian component location is highlighted in red, in panel c and d.

From these tests we can observe how, between the $\sim 2^{16}$ possible states composing the $N$-qubits wavefunction, the QAE algorithm is able to find, with a good probability, the bitstring corresponding to the right placement of the gaussian distributions in each mixture.

\section{Numerical simulation on CPUs for model systems with a VQA}
\label{num-sim-support}

For the VQA algorithm, a bayesian optimizer is employed to optimise the laser pulses and solve the posed problem. Several copies of the emulation of a given system are performed. We test the four different minimizers\footnote{Baysian Minimization functions are provided by the the \texttt{skopt} python library and used in the Pulser software} namely: i) Dummy, ii) Forest, iii) GBRT and iv) GP minimizer.

The parameters space size, denoted $m$, of $\Omega(t)$ and $\Delta(t)$ is  arbitrary. The larger is the 2$m$ dimensional parameter space, the larger is the flexibility left to the shape of the laser pulse. We tested $m$ values between $m = 3$ and $m = 25$. 

For each emulated system, 200 bayesian optimization cycles are performed. At the end of each cycle $k$, the reconstruction of the quatum state $\ket{\Psi^{\Omega,\Delta}_k}$ is emulated, so to evaluate the quality of the newly proposed set of laser parameters $(\Omega_k(t), \Delta_k(t))$, according to the cost function used by one of the proposed algorithms. The sampling procedure is emulated through Pulser and ideally corresponds to performing $200$ measures on $\ket{\Psi^{\Omega,\Delta}_k}$ on the real QPU. This number of repeated measures is a realistic parameter, affordable also on the real QPU. This corresponds in practice to sampling the frequency of appearence of the basis states in the $N$-qubits wavefunction
\begin{equation}
    \ket{\Psi^{\Omega,\Delta}_k} = \sum_{i=1}^{2^N} a_i\ket{e_i}.
\end{equation}

Assuming one, or several best configurations of excited atoms can be defined, namely the best basis state $\ket{e^{*}_{l}}$ in eq~\ref{eq:bs_best} in the main text, we measure the performance of the algorithm from the value of $||a_l^{*} ||^2$. In practice this is measured as the frequency of the state $\ket{e^{*}_{l}}$ in the sample obtained with 200 shots on the $N$-qubits quantum state applying the optimized lase pulse $(\Omega(t)^*,\Delta(t)^*)$.

For emulation purposes, we need to keep the size of the system under 15-18 qubits so to reduce the cost of solving numerically the TDSE, and allow us to perform a larger number of optimization cycles. 

\subsection{Insights on the quantum evolution}
A set of $M$-qubits wavefunctions is generated during the evolution, each represented as a superposition of qubits basis bitstrings, according to their respective frequencies. To do so, the quantum evolution process is ran (or emulated), starting from a state $\ket{\psi_k}$, to end up in the final state $\ket{\psi_{T}}$, where $T$ is the evolution period of the quantum state.
In other words, starting from $k=0$ and a prepared initial state $\ket{\psi_0}$, the Schrödinger equation is solved to yield $\ket{\psi_{k+1}}=SolveSchrodinger(\ket{ \psi_{k}})\;\forall k=0,\dots,(T-1)$ (cf. Algorithm~\ref{alg:HQC}).

The quantum evolution is repeated $n_c=200$ times, in as many Bayesian optimization cycles, yielding $\ket{\psi_T}_i$ for all $i=1,\dots,200$. For every generated $\ket{\psi_T}_i$ every accessible basis states $\ket{e}_i$ are sampled and their occurrences stored in $freq_{\ket{e}}$ yielding the aforementioned expected result. The sampling procedure is emulated through Pulser and  corresponds to performing $n_m:=200$ measures on the real QPU.

\subsection{Insights on the Baysian optimizer}
Concerning the proposed VQA, a Bayesian optimizer is used to guide the choice of the laser parameters. For each cycle of the Bayesian optimization, the best form of the functions $\Omega(t)$ and $\Delta(t)$ is searched. The quality of the explored functions shapes is measured by the distance between the reference density and the density reproduced by the qubits quantum state $\ket{\psi}$.

The parameters space size for the laser pulse optimization, namely $\Omega(t)$ and $\Delta(t)$, is itself an arbitrary parameter required by the algorithm. The two functions are described by a sum of elementary functions: $\sum_{i=1}^mf_m(t)$. The upper bound $m\in\mathbb{N}$ of the sum corresponds to the size of the parameters space.

\subsection{Numerical tests}

\begin{figure}[!htbp]
\begin{center}
\subfloat[]{\includegraphics[width=0.5\textwidth]{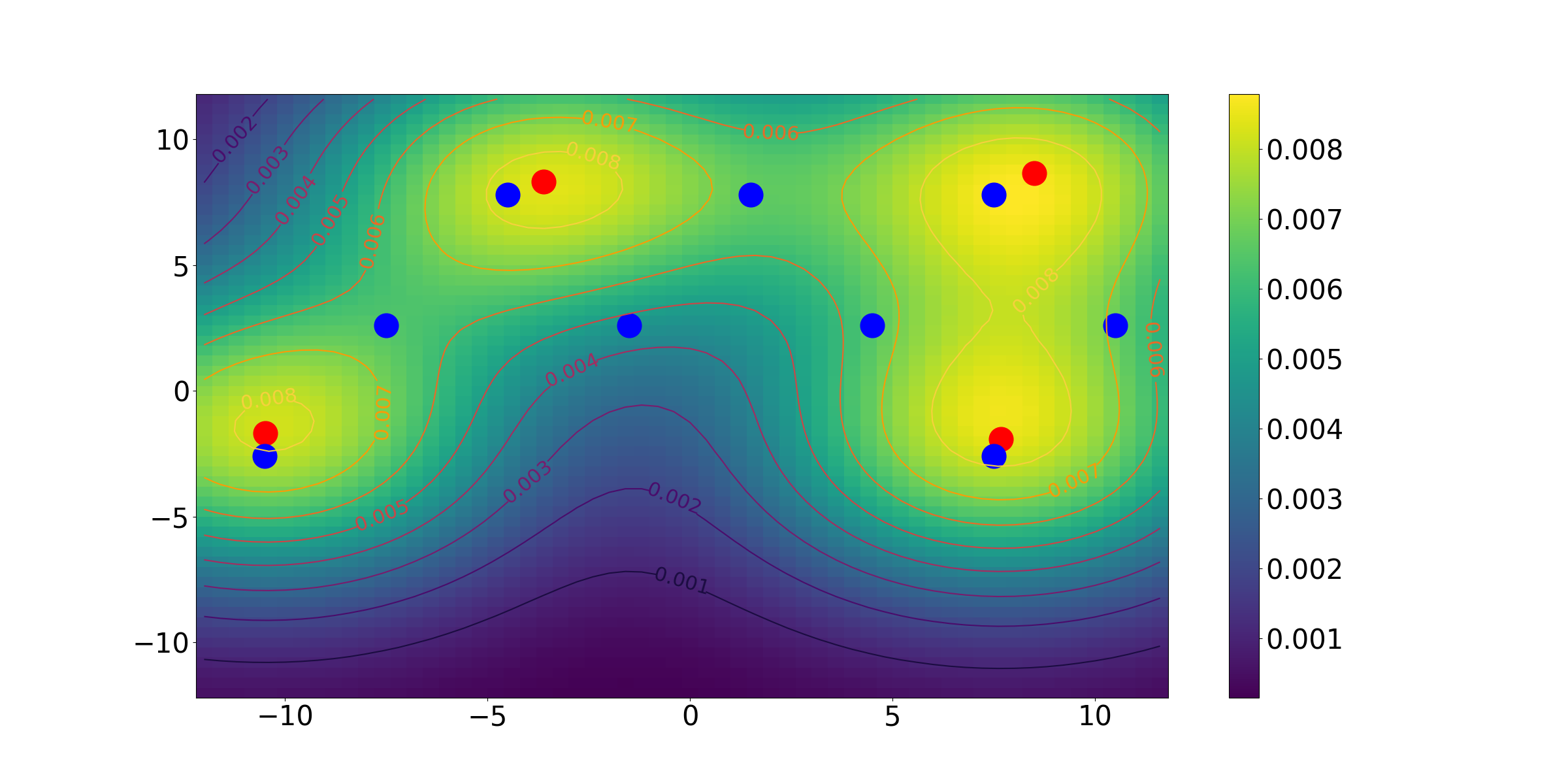}}
\subfloat[]{\includegraphics[width=0.5\textwidth]{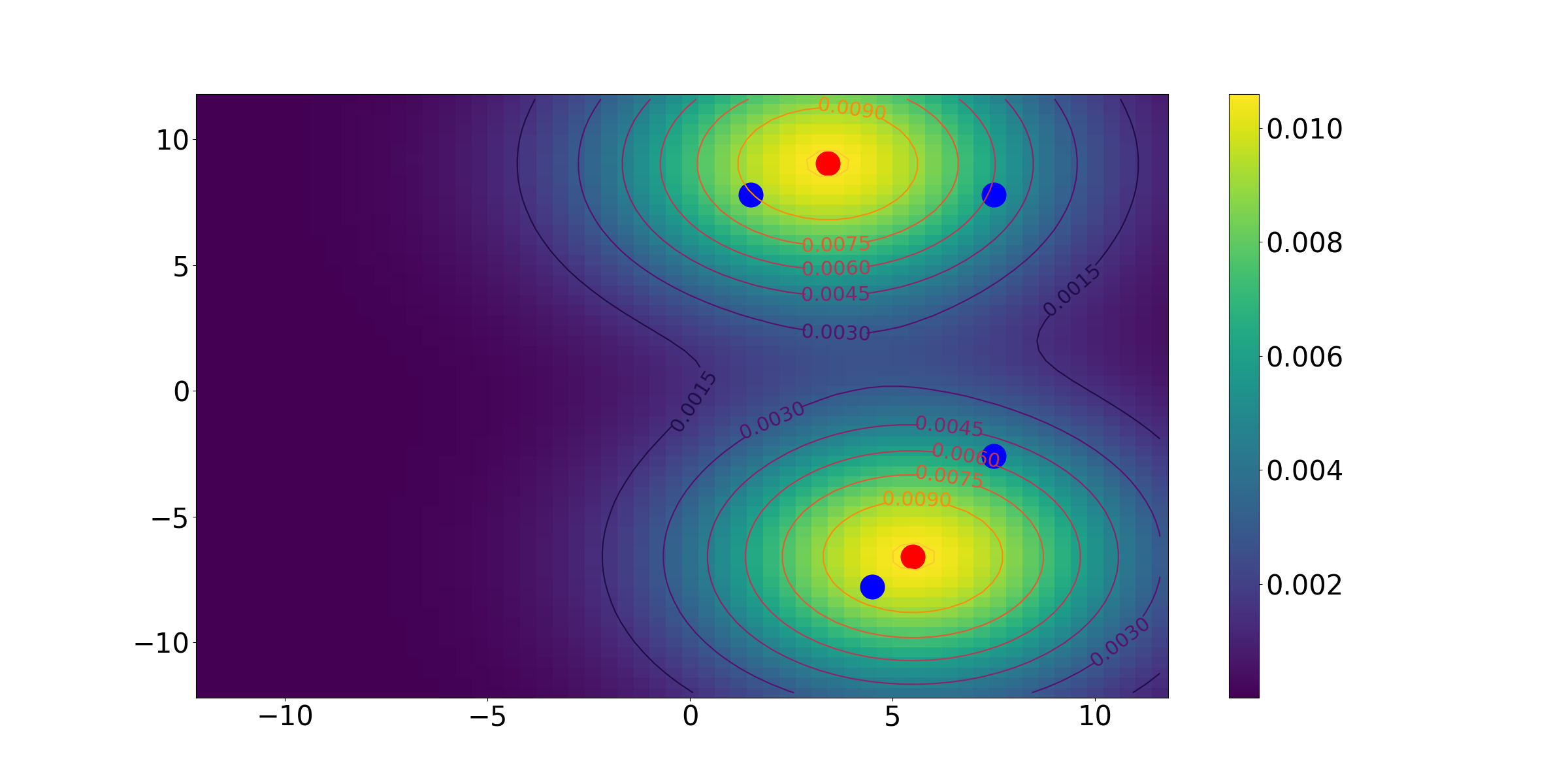}} \\
\subfloat[]{\includegraphics[width=0.5\textwidth]{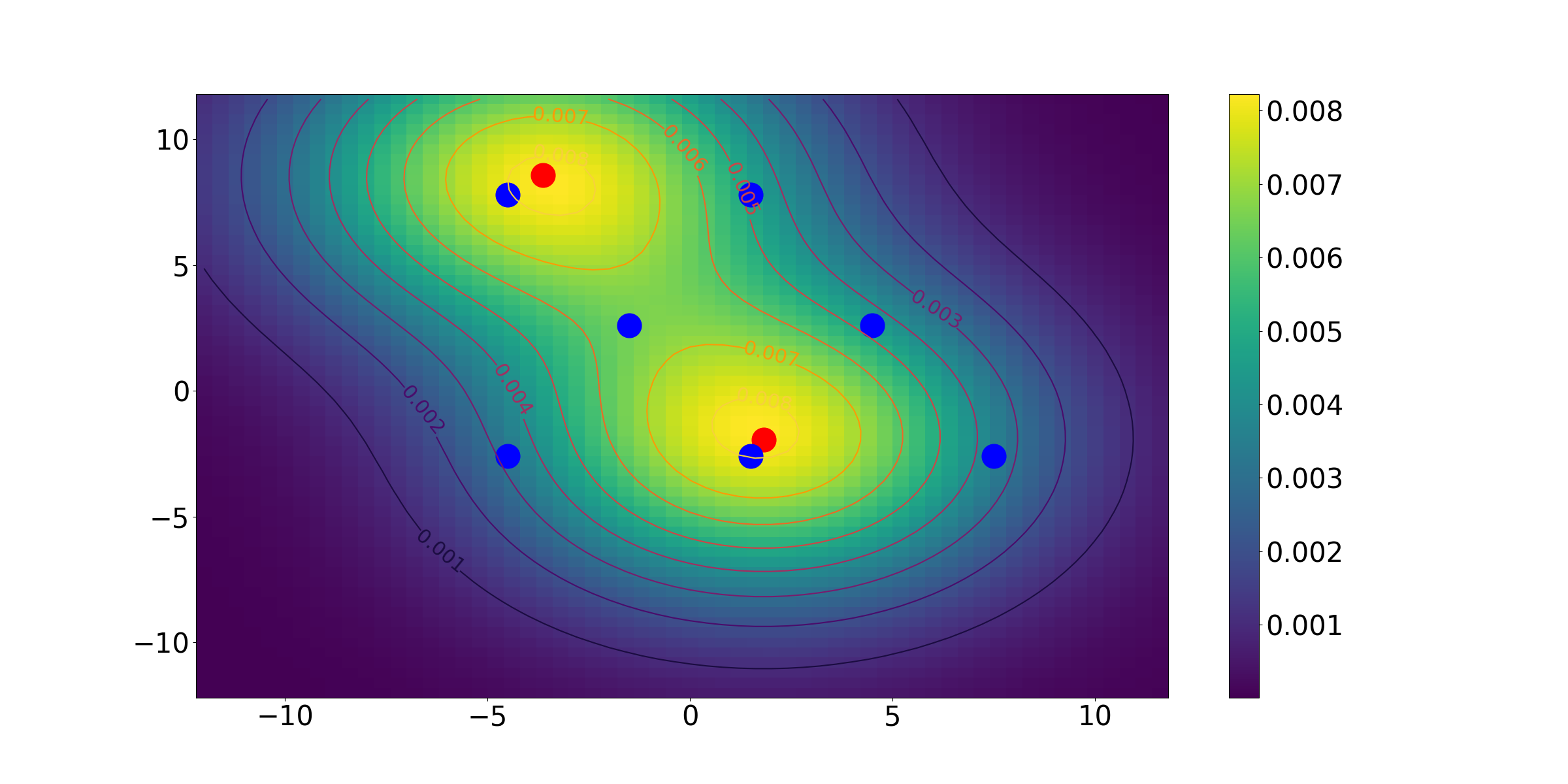}}
\end{center}
    \caption{Representation of the test synthetic sum of gaussian densities in 2D. The density maps, with level curves, are represented together with the qubits array (blue dots) built to address the location of the gaussian components. The centroid of each gaussian component is highlighted in red}
    \label{fig:2D_dens}
\end{figure}

To test the performances of the variational algorithm described in Algorithm~\ref{alg:HQC_IsG}, we used three simple 2D synthetic densities, reported in Figure~\ref{fig:2D_dens}. 

We have chosen on purpose simple synthetic densities to limit the number of qubits to employ, making possible to perform multiple times, during the optimization cycles, the numerically expensive emulations on classical CPUs. The Pulser emulator is used for this purpose, running the calculations on Atlas, the Qubit's hybrid CPU-GPU proprietary cluster. The best configuration of excited qubits is in these cases easy to identify: the set of excited qubits has to be composed by the qubits closer to the gaussians centroids, highlighted in red in Fig.~\ref{fig:exampleD}. A full cycle of optimization is performed for each test system, with the role of the QPU is taken by a classical solver for the time-dependent Shr{\"o}dinger equation.

For the second (VQA) proposed algorithm, the output of each emulation is the sampling of the optimized $N$-qubits state $\ket{\Psi^{\Omega^*,\Delta^*}}$, obtained after the hybrid quantum/classical optimization cycles. The optimizer employed employs Bayesian inference, introducing a stocastic behavior into the optimization procedure. Instead of performing a single optimization, reporting the final sampling of $\ket{\Psi^{\Omega^*,\Delta^*}}$, we perform several independent emulations for each test system. 

\begin{figure}[!htbp]
\begin{center}
\subfloat[]{\includegraphics[width=0.4\textwidth]{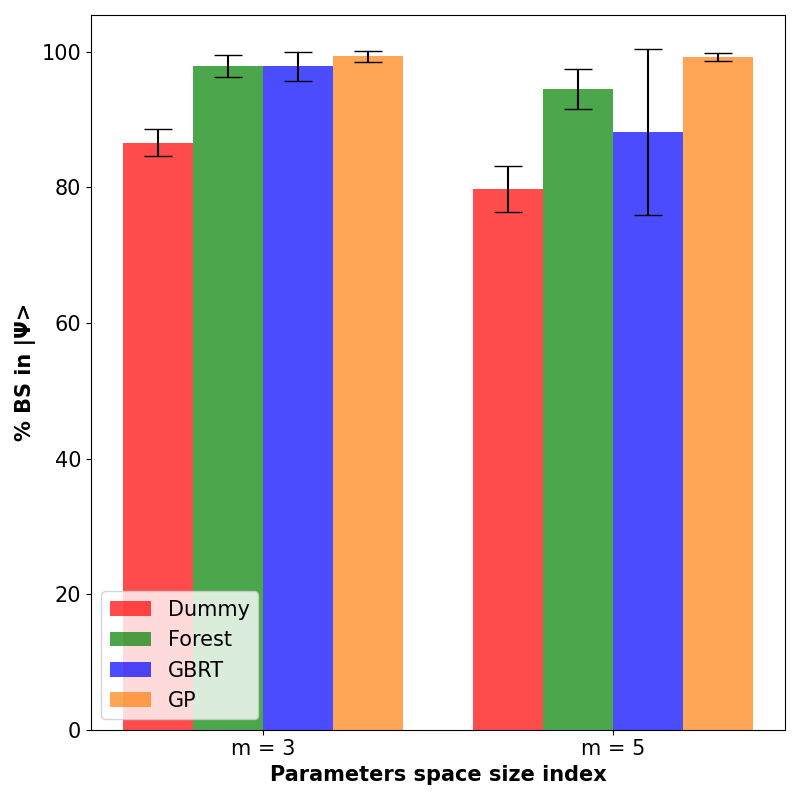}}
\subfloat[]{\includegraphics[width=0.4\textwidth]{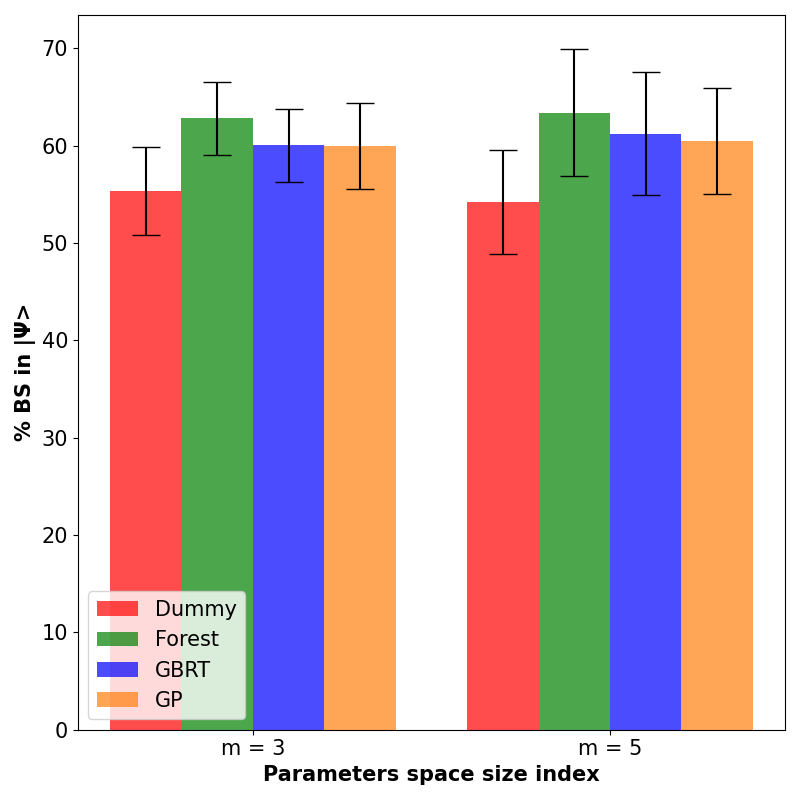}} \\
\subfloat[]{\includegraphics[width=0.4\textwidth]{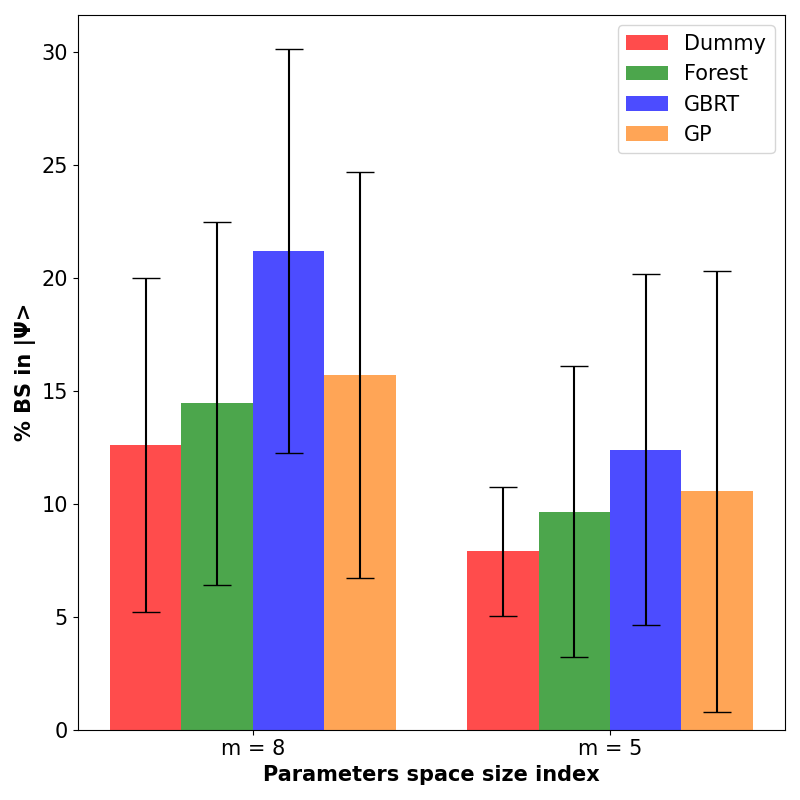}}
\subfloat[]{\includegraphics[width=0.4\textwidth]{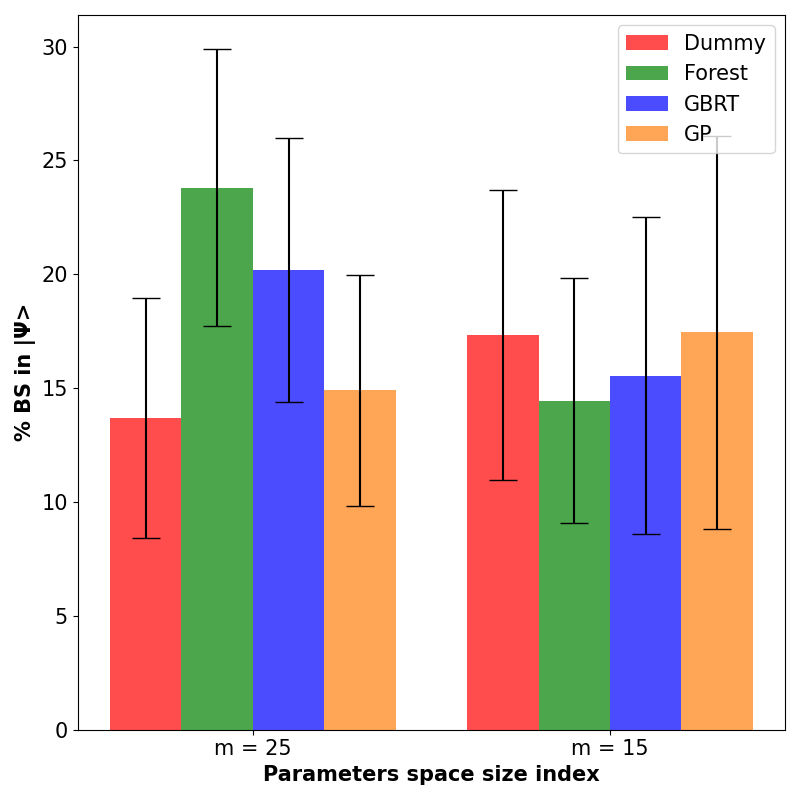}} 
\end{center}
    \caption{Population of the best qubit-string (BS), representing the solution for the three test densities; the population is reported as a percentage on the overall sampling of the optimized wave function $\ket{\Psi^{\Omega^*,\Delta^*}}$. Results from simulations using different algorithm depths are compared. The parameter m represents the size of the optimization space (number of points used to discretize the laser pulse shape). Higher the allowed parameter optimization is, the better the solution can be found by a suitable algorithm}
    \label{fig:2D_globalIsing}
\end{figure}

To have a global vision of the results, we report in Fig.~\ref{fig:2D_globalIsing} the averaged probability of only the best qubit-string (BS) $\ket{e^*}$, which best represents the reference density. For the four and two components gaussian densities (a and b in Fig.~\ref{fig:2D_globalIsing}), the BS represents between 100\% and 60\% of the total sample of the optimized state. Such an high percentage is also due to the fact that the actual solution of the problem corresponds to an MIS on the qubit array, which is the best scenario for Pasqal's QPU. Nevertheless, also the third case (c in Fig.~\ref{fig:2D_dens}), with a denser qubit distribution, shows a satisfactory accuracy, even if the contribution of $\ket{e^*}$ drops quite significantly. In this last case we show the beneficial effects of enlarging the allowed parameter space for the optimization, improving the level of detail for the laser pulse shaping (m parameter reported in all histograms).

\section{Numerical simulation on 3D-RISM 2D slices from MPU-I test system}

\begin{figure}[!htbp]
\begin{center}
\includegraphics[width=0.99\textwidth]{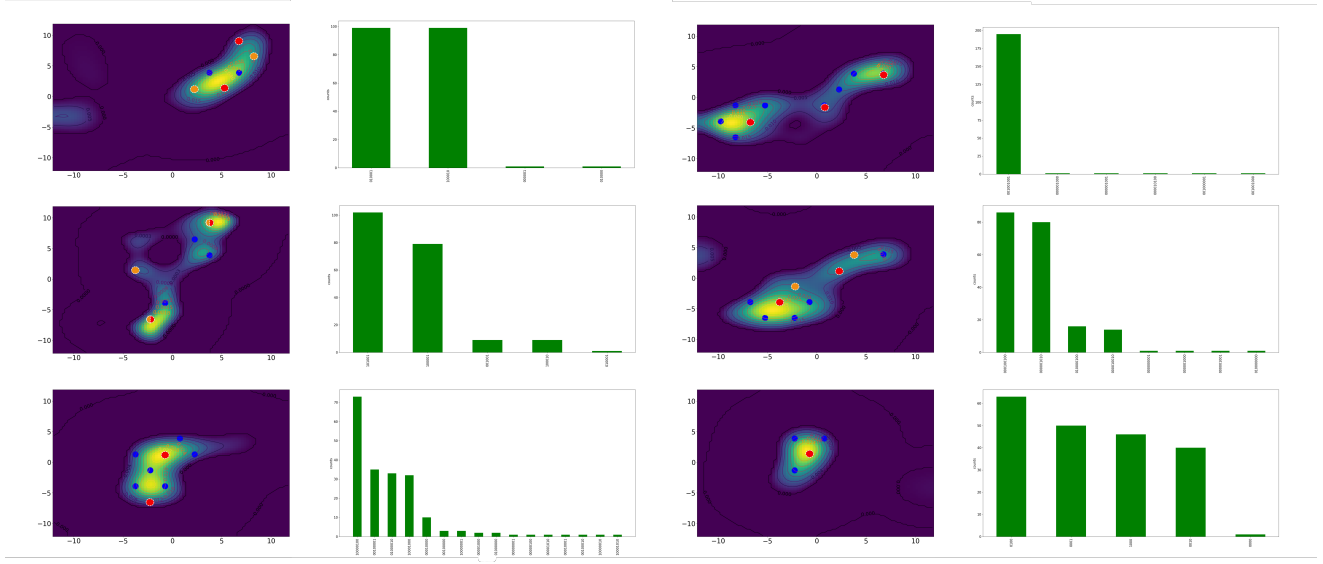}
\end{center}
    \caption{Representation of the test densities computed for the MUP-I protein, including the co-crystalized ligand, using the 3D-RISM method. We reported five 2D slices of the total 3D density, with level curves, on which the qubits array (blue dots) is built, to address the location of the signle gaussian components. The centroid of each gaussian component, found as an algorithm solution, is highlighted in red, for the highest probability bitstring, and orange, for the second highest probability bitstring.}
    \label{fig:3DRISM_slices_res}
\end{figure}

Here we report the emulation results, using the VQA, obtained for all the six 2D slices obtained from the 3D-RISM density computed from the MUP-I system, with co-crystallized ligand (2-sec-butyl-4,5-dihydrothiazole.
In Fig.~\ref{fig:3DRISM_slices_res} the six slices, and corresponding output hinstograms, are reported and compared.


\subsection*{A short digression on 3D-RISM}

Statistical mechanical theories, based on distribution functions and integral equations, can provide a rigorous framework for describing liquid phases, incorporating the granularity, packing, and specific molecular interactions without needing an explicit atoms-based representation of large ensembles of solvent molecules.

An example is the reference interaction site model (RISM)~\cite{chandler_optimized_1972SI,chandler_derivation_1973SI,beglov_integral_1997SI,perkyns_sitesite_1992,perkyns_dielectrically_1992,roy_chapter_2021,chandler_equilibrium_1974SI} proposed first by Chandler and Andersen in 1972, and 3D-RISM (the version of RISM extended to 3D)~\cite{cortis_three-dimensional_1997,kovalenko_three-dimensional_2003SI,kovalenko_self-consistent_1999,luchko_three-dimensional_2010,roy_chapter_2021}. 

3D-RISM represents a widely spread approach to treat the problem of solvation, as testified by the highly cited original works and by the implementation of the model in one of the standard software for molecular dynamics simulations, namely the Amber suite of programs.

The 3D-RISM theory provides probability density of all the solvent interaction sites $\gamma$ around a solute at position $\mathbf{r}$ as the product 
\begin{equation*}
    \rho_{\gamma}(\mathbf{r}) = \rho_{\gamma}g_{\gamma}(\mathbf{r})
\end{equation*}

of the average number density $\rho_{\gamma}$ in the bulk solution, and the
normalized density distribution $g_{\gamma}(\mathbf{r})$

The density distribution $g_{\gamma}(\mathbf{r})$ is the fundamental output of the model, with a site $\gamma$, corresponding to one of the physically different atoms which define the solvent model. 
High values of $g_{\gamma}(r)$ corresponds to regions with high solvent concentration, compared to the pure solvent bulk. One can interpret those high density regions as
location of stable water molecules, due to strong water-solute interactions.

In practice, 3D-RISM is an integral, site-site theory which needs as input the molecular structure of solute which one needs to solvate, and the solute and solvent models parameters, to compute the inter- and intra-molecular interactions between them.

A grid representation of the solvent around the solute, is used. The $g(r)$ function is evaluated on the grid, and each site of the solvent model will have a corresponding distribution value.

The 3D-RISM fundamental integral equation is the the solute-solvent total correlation function (TCF)~\cite{kovalenko_three-dimensional_2003SI,kovalenko_self-consistent_1999,luchko_three-dimensional_2010}

\begin{equation}
    h(r) = \sum_\alpha \int  c_\alpha^{UV}(r - r')\chi_{\alpha}^{VV}(r') dr^\prime
\end{equation}

for the solvent site $\gamma$. The superscripts ``U'' and ``V'' denote the solute and solvent species; the sum runs over the $\alpha$ solvent sites. $c_\alpha^{UV}$ is the 3D direct correlation function (DCF) for solvent site $\alpha$, which is proportional to the interaction potential between the solute and solvent site $u_\alpha^{UV}(r)/k_{B}T$, with opposite sign. $u_\alpha$ can be seen as the sum of pairwise solute-solvent, site-site potentials given from all the solute interaction sites located at frozen positions $R_i$.

The site-site susceptibility of the solvent is given by

\begin{equation}
\chi_{\alpha\gamma}^{VV}(r) = \omega_{\alpha\gamma}^{VV}(r) + \rho_\alpha h_{\alpha\gamma}^{VV}(r) 
\end{equation}

with $\omega_{\alpha\gamma}^{VV}$ the intramolecular correlation function,  representing the internal geometry of the solvent molecules, while $h_{\alpha\gamma}^{VV}$ is the site-site radial total correlation function of the pure solvent calculated from RISM~\cite{chandler_optimized_1972SI,beglov_integral_1997SI,perkyns_sitesite_1992,perkyns_dielectrically_1992}. A reference model for the solvent is chosen, and the susceptibility is computed accordingly. $\rho_\alpha$ is the solvent site bulk density.  

\begin{equation}
    h(r) = g(r) - 1
\end{equation}

is the relation which connects the main 3D-RISM integral equation to the desired solvent structure distribution function $g(r)$. 
Here, $h_\alpha$ and $c_\alpha$ are unknown. Therefore, in order to calculate $g$, the 3D-RISM algorithms uses an additional closure equation given by a Taylor expansion of order n:

\begin{equation}
    g_\alpha(r) = \left\{ \begin{array}{ll} \exp(t_\alpha(r)) & \text{if }  t_\alpha < 0 \\ \sum_{i=0}^{n} \frac{(t_\alpha(r))^i}{i!} & \text{if }  t_\alpha \geq 0 \end{array} \right.
\end{equation}

Where $t_\alpha(r) = -\frac{u_\alpha}{k_B T} + h_\alpha(r)-c_\alpha(r)$. When  $n=1$, this equation is called the Kovalenko-Hirata (KH) closure. When $n=\infty$, it is the so-called hypernetted $\lambda$-chain equation (HNC).

A lot of attention has been put in applying the model to molecular liquids~\cite{kinoshita_application_1996,roy_application_2020}, to study the solvation of
small molecules~\cite{truchon_cavity_2014,johnson_small_2016} or ion solvation~\cite{kovalenko_potentials_2000}, and even recently to larger biomolecules.~\cite{imai_ligand_2009,luchko_three-dimensional_2010,perkyns_protein_2010,sindhikara_analysis_2013}

To our knowledge, only recently 3D-RISM has been applied to study the hydration states of protein binding pockets~\cite{masters_efficient_2018,fusani_optimal_2018SI,nguyen_molecular_2019}, where classical computing technologies, based either on genetic algorithms (GAsol~\cite{masters_efficient_2018,fusani_optimal_2018SI}) and/or on molecular dynamics.~\cite{masters_efficient_2018,nguyen_molecular_2019}

\bibliography{SI_biblio.bib}